\pdfoutput=1

\documentclass[a4paper,fleqn,usenatbib]{mnras}

\usepackage{newtxtext} % newtxmath
\usepackage[T1]{fontenc}
\usepackage{ae,aecompl}

\usepackage[usenames]{color}

\newcommand{\dsct}{$\delta$~Sct }

\newcommand{\Kepler}{\textit{Kepler} }

\newcommand{\kms}{{\mathrm{km\,s^{-1}}}}

\usepackage{graphicx}	% Including figure files
\usepackage{amsmath}	% Advanced maths commands
\title[The unique pulsator KIC~5950759]{KIC~5950759: a high-amplitude \dsct star with amplitude and frequency modulation near the terminal age main sequence\thanks{Based on observations made with the WHT operated on the island of La Palma by the Isaac Newton Group in the Spanish Observatorio del Roque de los Muchachos of the Instituto de Astrof{\'i}sica de Canarias.}}

\author[D. M. Bowman et al.]{D. M. Bowman,$^{1}$\thanks{E-mail: dominic.bowman@kuleuven.be}
J. Hermans,$^{2,1}$
J. Daszy{\'n}ska-Daszkiewicz,$^{3}$
D. L. Holdsworth,$^{4}$
\newauthor
A. Tkachenko,$^{1}$
S. J. Murphy,$^{5, 6}$
B. Smalley,$^{7}$
D. W. Kurtz$^{8, 4}$
\\
$^{1}$ Institute of Astronomy, KU Leuven, Celestijnenlaan 200D, 3001 Leuven, Belgium \\
$^{2}$ Centre for mathematical Plasma Astrophysics, KU Leuven, 3001, Leuven, Belgium \\
$^{3}$ Astronomical Institute, University of Wroc{\l}aw, ul. Kopernika 11, 51-622 Wroc{\l}aw, Poland \\
$^{4}$ Jeremiah Horrocks Institute, University of Central Lancashire, Preston PR1 2HE, UK \\
$^{5}$ Sydney Institute for Astronomy (SIfA), School of Physics, University of Sydney, NSW 2006, Australia \\
$^{6}$ Stellar Astrophysics Centre, Department of Physics and Astronomy, Aarhus University, DK-8000 Aarhus C, Denmark \\
$^{7}$ Astrophysics Group, Lennard-Jones Laboratories, Keele University, Staffordshire ST5 5BG, UK \\
$^{8}$ Centre for Space Research, Physics Department, North-West University, Mahikeng 2745, South Africa \\
}

\date{Accepted 2021 April 16. Received 2021 April 13; in original form 2021 March 16}

\pubyear{2021}

\begin{document}
\label{firstpage}
\pagerange{\pageref{firstpage}--\pageref{lastpage}}
\maketitle

\begin{abstract}
Amongst the intermediate mass pulsating stars known as \dsct stars is a subset of high-amplitude and predominantly radial-mode pulsators known as high-amplitude \dsct (HADS) stars. From more than 2000 \dsct stars observed by the \Kepler space mission, only two HADS stars were detected. We investigate the more perplexing of these two HADS stars, KIC~5950759. We study its variability using ground- and space-based photometry, determine its atmospheric parameters from spectroscopy and perform asteroseismic modelling to constrain its mass and evolutionary stage. From spectroscopy, we find that KIC~5950759 is a metal-poor star, which is in agreement with the inferred metallicity needed to reproduce its pulsation mode frequencies from non-adiabatic pulsation models. Furthermore, we combine ground-based WASP and \Kepler space photometry, and measure a linear change in period of order $\dot{P}/P \simeq 10^{-6}$~yr$^{-1}$ for both the fundamental and first overtone radial modes across a time base of several years, which is at least two orders of magnitude larger than predicted by evolution models, and is the largest measured period change in a \dsct star to date. Our analysis indicates that KIC~5950759 is a metal-poor HADS star near the short-lived contraction phase and the terminal-age main sequence, with its sub-solar metallicity making it a candidate SX~Phe star. KIC~5950759 is a unique object amongst the thousands of known \dsct stars and warrants further study to ascertain why its pulsation modes are evolving remarkably faster than predicted by stellar evolution.
\end{abstract}


%%%%%%%%%%%%%%%%%%%%%%%%%%%%%%%%%%%%%%%%%%%%%%%%%%

%%%%%%%%%%%%%%%%% BODY OF PAPER %%%%%%%%%%%%%%%%%%

\section{Introduction}
\label{section: intro}

One of the goals of astrophysics is to improve models of stellar evolution by using constraints from observations. Understanding the physical processes inside a star, such as convection, rotation, angular momentum transport and mixing are of vital importance in improving stellar evolutionary models \citep{Maeder2000a, Maeder_rotation_BOOK}. This is especially true for intermediate- and high-mass stars since they are born with a convective core, and the theoretical uncertainties associated with convection and mixing in the core-envelope boundary region strongly influence their post-main sequence evolution \citep{Kippenhahn_BOOK}. Similarly, when such stars are close to depleting the hydrogen fuel in their cores, they enter an overall contraction phase to restore hydrostatic equilibrium, which results in a blue hook in the Hertzsprung--Russell (HR) diagram \citep{Iben1967a}. The location and morphology of this blue hook depends on the mass, metallicity, rotation rate and the properties of convection within the star. One of the most promising techniques to mitigate these uncertainties and ultimately improve the predictive power of stellar evolution models is using observations of stellar pulsations. In this technique, known as asteroseismology, the resonant pulsation frequencies of a star are used to constrain its interior properties from a quantitative comparison to stellar models containing different input physics \citep{ASTERO_BOOK, Aerts2021a}.

Amongst intermediate mass Population~{\sc I} stars of spectral type A and F is a group of pulsating stars known as \dsct stars, which have pulsation periods that range from $\sim$15~min to several hours \citep{Breger2000b, ASTERO_BOOK, Bowman2018a}. They are located at the intersection of the main sequence and the classical instability strip in the HR~diagram and have temperatures that range from approximately 7500 to 9500~K at the zero-age main sequence (ZAMS) and from approximately 6500 to 8500~K at the terminal-age main-sequence (TAMS; \citealt{Bowman2018a, Murphy2019a}). Thousands of \dsct stars are known based on ground-based observations and surveys \citep{Rod2000b, Jayasinghe2020c} and more recently space-based telescopes including \Kepler \citep{Borucki2010, Uytterhoeven2011, Balona2011g} and TESS \citep{Ricker2015, Antoci2019a}. The parameter space covered by \dsct stars in the HR~diagram is of great importance for testing stellar evolution models. They cover the transition region from slowly-rotating low-mass stars with radiative cores and thick convective envelopes ($M \leq 1.5$~M$_{\odot}$) to rapidly-rotating intermediate mass stars with convective cores and predominantly radiative envelopes ($M \geq 2.5$~M$_{\odot}$). This transition in stellar structure allows many different aspects of physics to be investigated, including pulsation, rotation, magnetic fields and chemical peculiarities (see \citealt{Murphy_PhD} for a review).

Pulsations in \dsct stars are typically excited by the opacity-driven heat engine $\kappa$ mechanism operating in the He~{\sc ii} ionization zone, which produces radial ($\ell=0$) and non-radial ($\ell > 0$) pressure (p) modes with pulsation periods between $\sim$15~min and several hours \citep{Breger2000b, Pamyat2000a}. Many \dsct stars are also observed to exhibit low-frequency variability caused by gravity (g) modes (see e.g. \citealt{Balona2011g}), which are not predicted to be excited in early-A main-sequence stars without a significant modification to stellar opacity tables \citep{Dupret2005a, Balona2015e}. Recent work on the excitation of pulsations in these stars has revealed that turbulent pressure is also important for exciting higher radial order p~modes \citep{Antoci2014b}, and that the efficiency of mixing and the helium abundance in the near-surface layers of \dsct stars plays an important role \citep{Antoci2019a, Murphy2020c}. 

Within the pulsator group of the \dsct stars, a small subset is observed to exhibit large pulsation amplitudes; these are known as high-amplitude \dsct (HADS) stars \citep{McNamara1997b, McNamara2000a}\footnote{\citet{McNamara1997b} credit the classification of HADS stars as a distinct group of pulsators to an unpublished Harvard university dissertation of H. Smith in 1955.}. The HADS stars are in general slow rotators with projected surface rotational velocities of $v\,\sin\,i \leq 40$~km\,s$^{-1}$, which is in contrast to the moderate and high rotation rates of main-sequence A stars \citep{Zorec2012, Niemczura2015}. In terms of their pulsation properties, HADS stars predominantly pulsate in the fundamental and/or first overtone radial p~mode \citep{Petersen1996, Breger2000a, McNamara2000a, Rod2000b, Derekas2009a}. The HADS stars were investigated by \citet{McNamara2000a}, who classified them as \dsct stars with large peak-to-peak light amplitudes (typically exceeding 0.3~mag) and pulsation periods between 1 and 6~hr. However, this historical classification is based only on inspection of the light curve, so it remains unclear why these stars are so rare and how they are distinct from their low amplitude \dsct star cousins. For example, \citet{Lee_Y_2008} estimated that less than 1~per~cent of pulsating A and F stars are HADS stars. Moreover, only two HADS stars were found within the sample of more than 10\,000 A and F stars observed by the \Kepler mission \citep{Bowman2016a, Balona2016b}. Some HADS stars have been shown to be members of multiple systems based on their measured radial velocities (e.g. \citealt{Derekas2009a}), although a complete census of binarity amongst HADS stars is lacking owing to their rarity.

In the literature HADS stars have been sometimes thought of as transition objects between the main sequence \dsct stars and the more-evolved classical (i.e. radial) pulsators, as they exhibit a tight period-luminosity relationship similar to Cepheid variables (see e.g. \citealt{Soszynski2008b, Poleski2010a}). On the other hand, a period-luminosity relationship for \dsct stars investigated by \citet{Ziaali2019a} contained considerably more scatter owing to these stars typically having both radial and non-radial modes with comparable amplitudes that exhibit long-term amplitude modulation \citep{Bowman2016a}. Because of their rarity, there have only been a few detailed studies of HADS stars in the literature, which aimed to determine if these stars are distinct from \dsct stars. For example, \citet{Petersen1996} claimed that HADS stars are able to pulsate at significantly higher amplitudes because they are in a post-main sequence stage of stellar evolution. On the other hand, \citet{Breger2000a} conjectured that the slow rotation in these stars facilitates the high amplitude pulsations, which are not able to be excited in rapidly rotating \dsct stars because of the deformation from spherical symmetry.

The Population~{\sc II} counterparts of \dsct stars are known as SX~Phe stars, named after the first such star to show high-amplitude radial pulsations \citep{Frolov1984a}, although such high-amplitude pulsation has been known since \citet{Goodricke1783}. SX~Phe stars are also located within the classical instability strip in the HR~diagram, have similar pulsation frequencies to \dsct stars, and dominant radial pulsation modes with relatively large amplitudes like HADS stars \citep{Balona2012e}. However, they sometimes have sub-solar metallicities and large spatial motions \citep{McNamara2000a, Nemec2017a}. Interestingly, SX~Phe stars are known to exhibit period (i.e. pulsation frequency) changes when monitored over several years \citep{Coates1982b, Thompson_K_1991, Rod2007d, Yang_X_2012a, Murphy2013b}, and yet these changes are sometimes observed as sudden jumps in period rather than smooth variation over time (e.g. \citealt{Rod1995d}). Therefore, it remains unclear if the period changes and associated amplitude modulation of pulsation modes are directly caused by stellar evolution. Recently, \citet{Daszy2020c} performed detailed seismic modelling of the prototype of this class, SX~Phe, constrained its fundamental parameters and confirmed its post-main sequence evolutionary phase thanks to high-precision photometry assembled by the TESS mission.

In this paper, we present the study of the pulsating star KIC~5950759, which is one of only two HADS stars observed by the \Kepler space telescope during its nominal 4-yr mission. In section~\ref{section: photometry} we use \Kepler and WASP photometry to extract its pulsation mode frequencies and measure their long-term variability. In section~\ref{section: spectroscopy}, we present our spectroscopic analysis of KIC~5950759, and in section~\ref{section: gaia} we use Gaia parallaxes to calculate its luminosity. We perform forward asteroseismic modelling of KIC~5950759 in section~\ref{section: modelling}, and we conclude in section~\ref{section: conclusions}.

%%%%%%%%%%%%%%%%%%%%%%%%%%%%%%%%%%%%%%%%%%%%%%%%%%

\section{Photometry}
\label{section: photometry}

KIC~5950759 was first identified as a variable star with a dominant period of 0.0703176~d by \citet{Pigulski2009} using ASAS data. Later, it was identified as one of only two HADS stars in the nominal field of view of the \Kepler mission \citep{Bowman2016a, Balona2016b}, which satisfied the criterion of having large peak-to-peak light amplitude variations typical of such stars \citep{McNamara2000a}. The parameters of KIC~5950759 from the \Kepler Input Catalogue (KIC; \citealt{Brown2011}) and the revised values from \citet{Huber2014}, which are similar to the updated values of \citet{Mathur2017a}, are provided in Table~\ref{table: parameters}.

% - - - - - % - - - - - % - - - - - % - - - - - %
\begin{table}
\centering
\caption{Stellar parameters of KIC~5950759 from photometric catalogues and spectroscopy from this work (see text for details), including the effective temperature ($T_{\rm eff}$), surface gravity ($\log\,g$) and metallicity ([M/H]).}
\begin{tabular}{l l l l}
\hline
\multicolumn{1}{c}{T$_{\rm eff}$} & \multicolumn{1}{c}{$\log\,g$} & \multicolumn{1}{c}{[M/H]} &  \multicolumn{1}{c}{Reference} \\
\multicolumn{1}{c}{(K)} & \multicolumn{1}{c}{} & \multicolumn{1}{c}{(dex)} &  \multicolumn{1}{c}{} \\
\hline
$7840 \pm 300$	&	$4.03 \pm 0.25$	&	$-0.07 \pm 0.25$	&	\citet{Brown2011} 	\\
$8070 \pm 280$	&	$4.03 \pm 0.40$	&	$-0.07 \pm 0.30$	&	\citet{Huber2014}	\\
\hline
$7340 \pm 220$	&	$2.9 \pm 1.3$	&	$-1.18 \pm 0.30$	&	this work	 \\
$7470 \pm 255$	&	$3.6 \pm 1.2$	&	$-0.65 \pm 0.45$	&	this work 	\\
\hline
\end{tabular}
\label{table: parameters}
\end{table}
% - - - - - % - - - - - % - - - - - % - - - - - %

%	%	%	%	%	%	%	%	%	%
	
	\subsection{\Kepler Photometry}
	\label{subsection: Kepler}

	In this work, light curves from the multi-scale Maximum A Priori Pre-Search Data Conditioning (msMAP PDC) pipeline developed by the \Kepler Science Office \citep{Stumpe2012, Smith_J_2012} are used, which are publicly available from the Mikulski Archive for Space Telescopes (MAST)\footnote{\url{http://archive.stsci.edu/kepler/}}. KIC~5950759 was observed between Q1 and Q17 of the nominal \Kepler mission in its long cadence (LC; 29.5~min) mode, for a total of 1460~d. It was also observed in the first month of Q4 in short cadence (SC; 59~s) mode for a total of 31.1~d. The effective Nyquist frequencies of the LC and SC data are 24.47 and 727.35~d$^{-1}$, respectively. The \Kepler light curves were converted into stellar magnitudes and long-period instrumental systematics were removed by means of subtracting a low-order polynomial from each LC data quarter.
	
	\begin{figure}
	\centering
	\includegraphics[width=0.99\columnwidth]{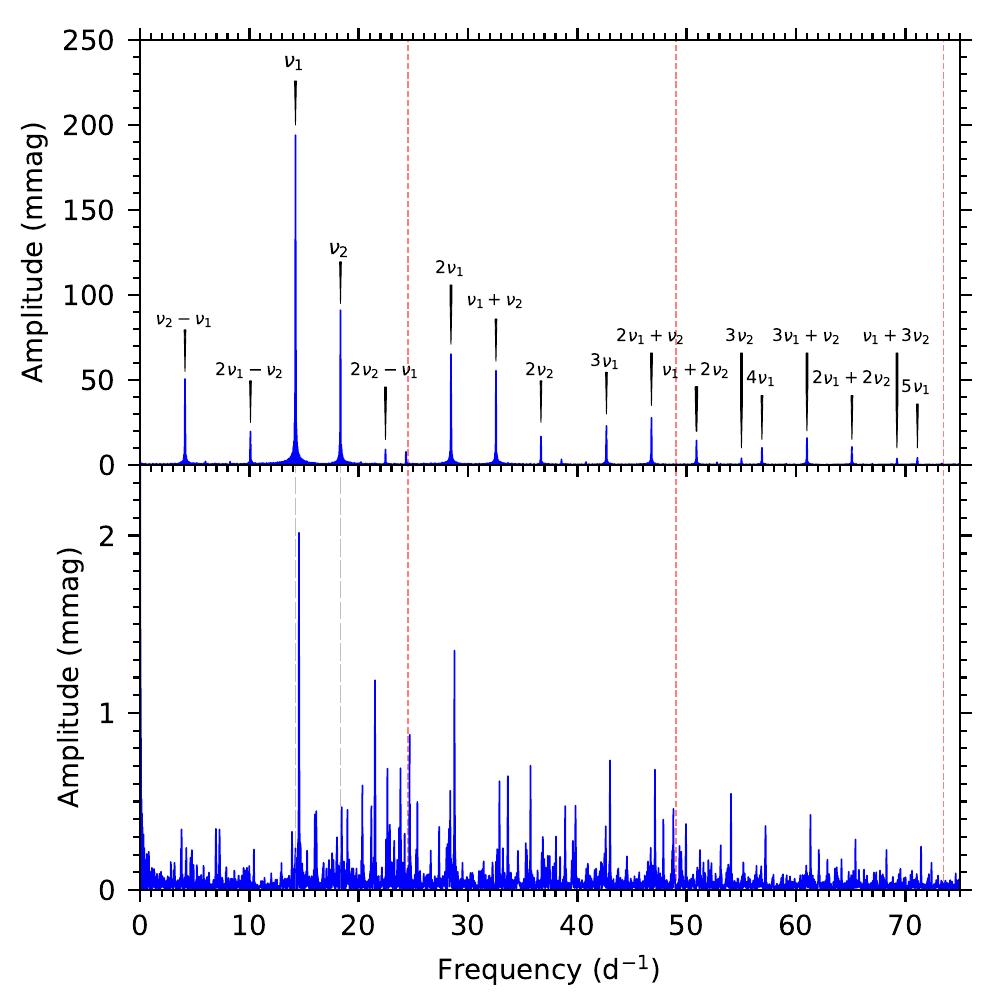}
	\caption{Top panel: the amplitude spectrum calculated from the one month of SC \Kepler data for KIC~5950759, in which the two dominant pulsation modes, $\nu_1$ and $\nu_2$, and their high-amplitude combination frequencies are labelled. Bottom panel: the residual amplitude spectrum of the same SC \Kepler data after $\nu_1$ and $\nu_2$ and all their significant combination frequencies have been removed by pre-whitening. The long-dashed grey lines denote the location of $\nu_1$ and $\nu_2$ in the bottom panel, and the short-dashed red lines indicate integer multiples of the \Kepler LC Nyquist frequency to demonstrate that considerable variance remains beyond the \Kepler LC Nyquist frequency.}
	\label{figure: SC FT}
	\end{figure}
	
	From the LC and SC light curves amplitude spectra were calculated by means of a discrete Fourier transform \citep{Kurtz1985b}. The amplitude spectrum of the single month of SC \Kepler data is shown in the top panel of Fig.~\ref{figure: SC FT}, in which the two high-amplitude pulsation modes, $\nu_1$ and $\nu_2$, and their dominant combination frequencies are labelled. The two high-amplitude modes have a period (and frequency) ratio of $0.77554$ identifying them as the fundamental and first overtone radial modes \citep{Breger2000b}. In \dsct stars with such high-amplitude (radial) pulsation modes, it is typical to observe many high-order harmonics and combination frequencies --- i.e. frequencies that can be identified using $n\nu_1 \pm m\nu_2$, in which $n$ and $m$ are integers \citep{Papics2012b, Kurtz2015b, Balona2016b}. The physical causes of combination frequencies include the stellar medium not responding linearly to a pulsation wave, or that the flux variability caused by waves is not a linear transformation of the temperature variability (see e.g. \citealt{Brickhill1992a}). These phenomena are generally grouped together into what is called a non-linear distortion model (e.g. \citealt{Degroote2009a, Bowman_PhD}). On the other hand, some combination frequencies represent resonantly excited pulsation modes that are the result of direct mode coupling (see e.g. \citealt{Breger2014, Bowman2016a}). 
	
% - - - - - % - - - - - % - - - - - % - - - - - %
\begin{table}
\centering
\caption{Frequency, amplitude, and phase of the two dominant pulsation modes, $\nu_1$ and $\nu_2$, in KIC~5950759, which were obtained using a non-linear least-squares fit to the 4-yr LC \Kepler light curve, and the 31-d SC light curve. Pulsation phases were calculated with respect to the time zero-point $t_0 = 2455688.770$~BJD.}
\begin{tabular}{r r c r c}
\hline
\multicolumn{1}{c}{} & \multicolumn{1}{c}{} & \multicolumn{1}{c}{Frequency} & \multicolumn{1}{c}{Amplitude} & \multicolumn{1}{c}{Phase} \\
\multicolumn{1}{c}{} & \multicolumn{1}{c}{} & \multicolumn{1}{c}{(d$^{-1}$)} & \multicolumn{1}{c}{(mmag)} & \multicolumn{1}{c}{(rad)} \\
\hline
LC	&	$\nu_1$	&	$14.2213938 \pm 0.0000009$	&	$167.7 \pm 0.3$	&	$1.685 \pm 0.002$	\\
	&	$\nu_2$	&	$18.3372939 \pm 0.0000015$	&	$74.1 \pm 0.3$		&	$1.178 \pm 0.004$	\\
\hline
SC	&	$\nu_1$	&	$14.22137 \pm 0.00005$	&	$194.1 \pm 0.5$	&	$1.6 \pm 0.1$	\\
	&	$\nu_2$	&	$18.33725 \pm 0.00010$	&	$91.2 \pm 0.5$		&	$1.0 \pm 0.3$	\\
\hline
\end{tabular}
\label{table: freqs}
\end{table}
% - - - - - % - - - - - % - - - - - % - - - - - %

	A detailed frequency analysis of both the LC and SC \Kepler data of KIC~5950759 was originally performed by \citet{Bowman_BOOK}, which focused on the two radial modes and their combination frequencies. Since the much longer 4-yr LC \Kepler data set provides significantly better frequency resolution and precision, \citet{Bowman_BOOK} extracted and optimised the parameters of the two dominant pulsation modes, $\nu_1$ and $\nu_2$, using a multi-frequency non-linear least-squares fit to the LC \Kepler light curve with the equation:
	
	\begin{equation}
	\Delta\,m = \sum_{i} A_i\,\cos\left(2\pi\nu_i\left(t-t_0\right)+\phi_i\right) ~ ,
	\label{equation: cosine}
	\end{equation}
	
	\noindent where $A_i$ is the amplitude, $\nu_i$ is the frequency, $\phi_i$ is the phase, $t$ is the time with respect to the zero-point $t_0 = 2455688.770$~BJD (i.e. the midpoint of the 4-yr \Kepler light curve) for $i$ frequencies. The resultant parameters for $\nu_1$ and $\nu_2$ are provided in Table~\ref{table: freqs}. 
	
	\citet{Bowman_BOOK} also performed a similar optimisation of the two dominant pulsation modes using Eq.~\ref{equation: cosine} and a non-linear least-squares fit to the SC \Kepler data of KIC~5950759. Since the SC \Kepler data has a much higher Nyquist frequency, \citet{Bowman_BOOK} calculated the expected frequencies of combination frequencies of $\nu_1$ and $\nu_2$ up to order of 20 (i.e. $n\nu_1 \pm m\nu_2$, where $n,m \in [0,20]$), and optimised their amplitudes and phases using a multi-frequency linear least-squares fit, and subtracted them from the SC \Kepler light curve to reveal any additional independent pulsation modes. Such high harmonic and combination orders are necessary for KIC~5950759 given the high amplitudes of the parent modes (i.e. $\nu_1$ and $\nu_2$) and the pristine quality of \Kepler data. For example, the twentieth harmonic of $\nu_1$ (i.e. $20\nu_1$) is significant in the SC data of KIC~5950759 with an amplitude of approximately 10~$\mu$mag. By using the approach of calculating the expected combination frequencies, any variance coinciding with the expected frequencies of combinations is removed from the light curve by means of subtracting the multi-frequency linear least-squares fit solution. The amplitude spectra of the SC \Kepler light curve up to 75~d$^{-1}$ both before and after removing $\nu_1$, $\nu_2$ and all their combination frequencies are shown in the top and bottom panels of Fig.~\ref{figure: SC FT}, respectively, which clearly demonstrates additional variance in KIC~5950759.
		
	\citet{Yang_T_2018b} later performed an independent frequency analysis of the available \Kepler data of KIC~5950759. They confirmed the previous characterisation of $\nu_1$ and $\nu_2$ as radial modes and investigated a detected weak modulation effect present in KIC~5950759, which resulted in equally-spaced frequency multiplets around the dominant pulsation modes. The separation between the frequencies of these multiplets, $0.31923$~d$^{-1}$, was also detected as a significant frequency in the amplitude spectrum. \citet{Yang_T_2018b} concluded that these multiplets could be caused by amplitude modulation of the radial modes from the rotation of the star\footnote{To be clear, this does not mean rotational splitting of radial modes, since this is not possible, but rather means that the pulsation amplitudes of the radial modes varies during the rotation phase causing equally-spaced multiplets in the amplitude spectrum.}. Such a modulation mechanism acting on radial modes has been found in some \dsct stars (e.g. \citealt{Breger2011}), but its physical cause remains difficult to interpret. The inferred rotation frequency from the multiplets, $0.31925$~d$^{-1}$, would indicate a slow rotation rate, which is expected for HADS stars. For a discussion of this modulation effect, we refer the reader to \citet{Yang_T_2018b}.
	
	More recently, \citet{Hermans_MSc} re-visited the frequency analysis of KIC~5950759 for the specific purpose of identifying additional non-radial modes. The same LC and SC light curves studied by \citet{Bowman_BOOK} were analysed by \citet{Hermans_MSc} for consistency and comparison purposes. The LC and SC data were subjected to iterative pre-whitening, and modelled as a linear superposition of modes, as is typical for pulsating stars (see e.g. \citealt{Degroote2009a, Papics2012a, VanReeth2015a, Antoci2019a}). During each iteration, the frequency of the highest amplitude peak is determined, optimised to the light curve using Eq.~(\ref{equation: cosine}), which yields the optimised frequency, amplitude and phase. The resulting cosinusoid model is then subtracted from the light curve and the residual time series is used as input in the next iteration. The standard stop criterion of an amplitude signal-to-noise (S/N) criterion of four \citep{Breger1993b} was used by \citet{Hermans_MSc} to extract all significant frequencies, in which the noise was calculated as the median amplitude in a window of 1~d$^{-1}$ centred on the extracted frequency at each iteration. 
	
	A total of 150 significant frequencies were identified in the LC \Kepler data in the frequency interval of $0 \leq \nu \leq 120$~d$^{-1}$, and 106 significant frequencies in the SC \Kepler data of KIC~5950759 in the frequency interval $0 \leq \nu \leq 200$~d$^{-1}$ by \citet{Hermans_MSc}. Although pulsation modes with frequencies above 60~d$^{-1}$ are rare in \dsct stars \citep{Bowman2018a}, the large frequency range was chosen to ensure that high frequency non-radial modes were not excluded. From these lists of extracted significant frequencies, spurious frequencies were filtered using the resolution criterion proposed by \citet{Loumos1978}, such that the minimal frequency separation between independent close frequencies, $\nu_x$ and $\nu_y$, satisfied
	
	\begin{equation}
	| \nu_x - \nu_y | \geq \frac{1.5}{\Delta\,T} ~ ,
	\label{equation: resolution}
	\end{equation}
	
	\noindent where $\Delta\,T$ is the length of the data set. The highest amplitude peak was kept during the filtering for spurious frequencies. Combination frequencies were identified by searching for linear sum and difference frequencies, $n\nu_i \pm m\nu_j$, with the \citet{Loumos1978} criterion as a tolerance and assuming that the highest-amplitude peaks within a combination family are the real pulsation mode frequencies \citep{Kurtz2015b}. Although the \citet{Loumos1978} criterion is not a formal frequency precision (see e.g. \citealt{Lares-Martiz2020a}), it does act as a conservative metric for identifying combination frequencies given the added uncertainties introduced in the form of spurious frequencies from an imperfect linear least-squares pre-whitening methodology (see e.g. \citealt{Degroote2009a, Papics2012a, VanReeth2015a, Antoci2019a}). Finally, Nyquist alias frequencies were identified by cross-matching the resultant list of filtered frequencies in the LC and SC analysis, and by applying the super-Nyquist technique \citep{Murphy2013a} to the LC \Kepler data.	
	
	In total, 12 and 13 independent pulsation mode frequencies (i.e. those that cannot be explained as spurious close frequencies, combination frequencies, or Nyquist aliases) were identified from the LC and SC data in addition to $\nu_1$ and $\nu_2$ (cf. Table~\ref{table: freqs}), which are given in Tables~\ref{table: residual LC list} and \ref{table: residual SC list}, respectively. Therefore, in addition to the two high-amplitude radial pulsation modes, $\nu_1$ and $\nu_2$, KIC~5950759 reveals multiple additional non-radial pulsation modes. Unfortunately, mode identification is not possible for these non-radial modes owing to their lack of regularity in the extracted frequencies. We also note that the same frequency multiplets found by \citet{Yang_T_2018b} centred on the radial mode frequencies are present in both LC and SC data, but have much lower S/N values in the SC data. For example, the modulation frequency, $0.31925 \pm 0.00004$~d$^{-1}$, is only significant in the LC data because of the lower local noise level.

%	%	%	%	%	%	%	%	%	%
	
	\subsection{WASP photometry}
	\label{subsection: WASP}
	
	Given the extremely high frequency precision provided by its high-amplitude radial modes using \Kepler data, \citet{Bowman_PhD} identified KIC~5950759 as an excellent candidate to probe pulsation period changes. Moreover, ground-based photometric surveys prior to and during the \Kepler mission have also been able to monitor its high-amplitude pulsation modes. A naturally complementary photometric survey to our analysis of the \Kepler light curve of KIC~5950759 is the Wide Angle Search for Planets (WASP) project, which was a wide-field and ground-based photometric survey that aimed to find transiting exoplanets \citep{Pollacco2006, Butters2010}. The project had two observing sites, one at the Observatorio del Roque de los Muchachos on La Palma and the other at the Sutherland Station of the South African Astronomical Observatory (SAAO). The pixel size of each instrument is approximately 14~arcsec with observations made using broad-band filters of $400-700$~nm, which are slightly more sensitive to bluer wavelengths compared to the \Kepler passband. WASP data are corrected for extinction, the colour-response of the instrument, the zero-point, and instrumental systematics using the {\sc sysrem} algorithm \citep{Tamuz2005}. The observing strategy of WASP provided two consecutive 30-s exposures at a given pointing, before moving to the next available field. Each field was typically revisited every 10 min on a given night, which produces a high effective Nyquist frequency \citep{Holdsworth_PhD}.
	
	\begin{figure*}
	\centering
	\includegraphics[width=0.95\textwidth]{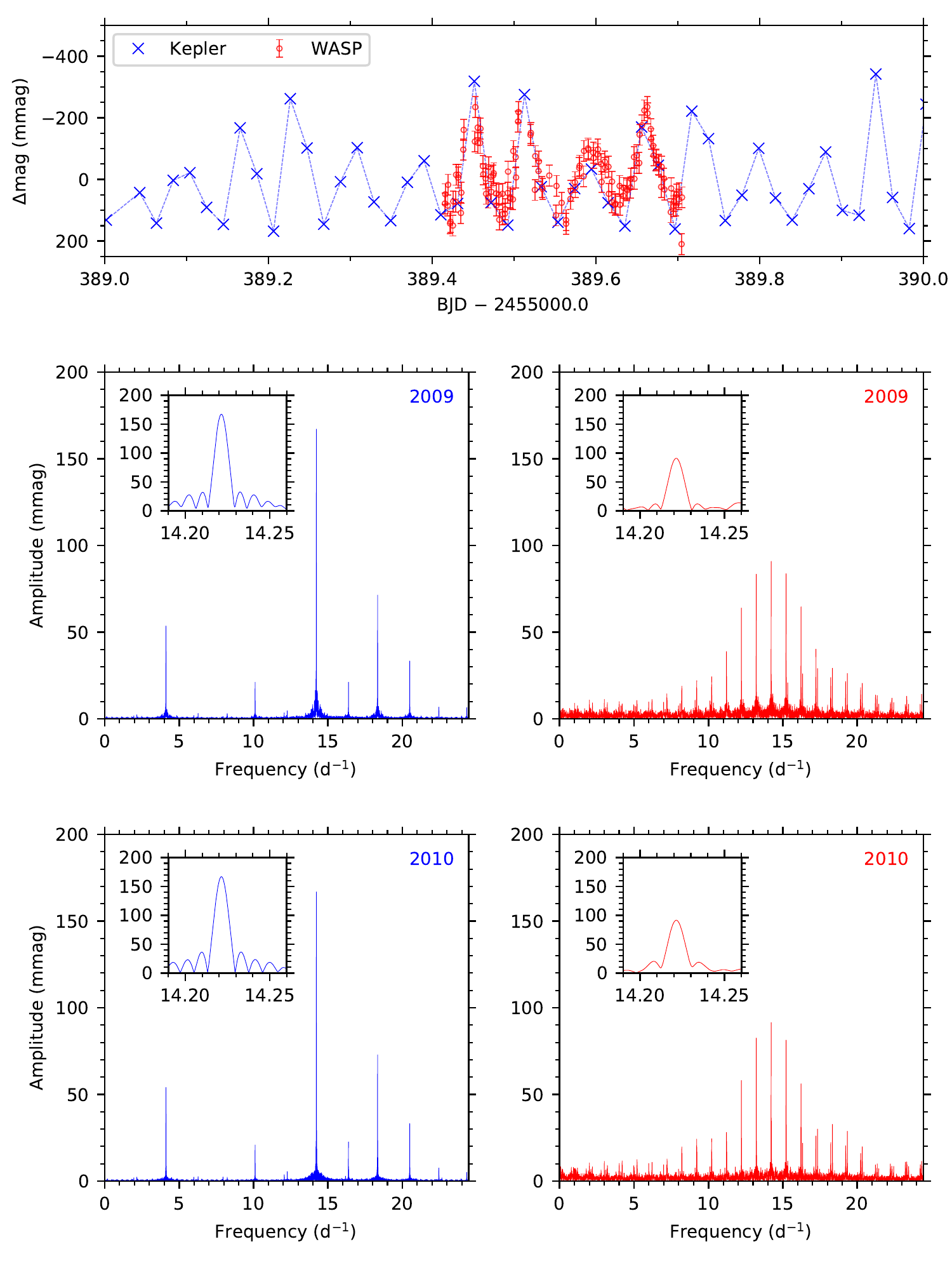}
	\caption{The top panel is a 1-d segment of the \Kepler and WASP light curves of KIC~5950759 shown as blue crosses and red circles, respectively. The lower panels show amplitude spectra of KIC~5950759 labelled by their respective observing season, with the left panels denoting \Kepler data in blue and the right panels denoting WASP data in red. In each panel, a sub-plot of the dominant pulsation mode frequency, $\nu_1$, is shown. The lower amplitudes of the pulsation modes in the WASP data is primarily caused by flux dilution in the relatively larger WASP pixels (see text for details).}
	\label{figure: grid}
	\end{figure*}
	
	The WASP survey observed KIC~5950759 for three seasons in 2007, 2009 and 2010, with light curves spanning 65.8, 127.9 and 122.9~d, respectively. Although the photometric precision and number of data points are not as high in each WASP light curve compared to the LC \Kepler data, the high-amplitude pulsation modes in KIC~5950759 are easily detectable. Moreover, during 2009 and 2010, data collection by both WASP and \Kepler were concurrent. This has great advantages and allows us to use the WASP data to probe the variability in KIC~5950759 prior to the launch of the \Kepler space telescope, as previously demonstrated using the \dsct star KIC~7106205 by \citet{Bowman2015a}. To compare simultaneous WASP and LC \Kepler observations of KIC~5950759 in 2009 and 2010, the \Kepler data were truncated to have the same length as the individual WASP seasons. Also, the time stamps of the WASP data, which were originally in HJD~(UTC), were converted to BJD~(TDB) using the \texttt{PYTHON} module \texttt{ASTROPY} \citep{Astropy_2013, Astropy_2018} to be fully consistent in both time scale and format with the \Kepler data. An example of the concurrent LC \Kepler and WASP light curves for KIC~5950759 in 2010 are shown in the top panel of Fig.~\ref{figure: grid}. The lower panels in Fig.~\ref{figure: grid} show the amplitude spectra of the WASP data (right-hand panels) and the truncated \Kepler data (left-hand panels). 
	
	Because of the discrete and finite sampling of an instrument's integration time, a star's pulsation modes are suppressed in amplitude because of an effect known as apodization or the amplitude suppression function \citep{Bowman2015a}. The observed pulsation amplitudes can be corrected using
	
	\begin{equation}
	A = A_0~{\rm sinc}\left(\frac{\pi}{n}\right) = A_0~{\rm sinc}\left(\frac{\pi \nu}{\nu_{\rm samp}}\right) ~ ,
	\label{equation: integration}
	\end{equation}
	
	\noindent where $A$ and $A_0$ are the observed and intrinsic pulsation mode amplitudes, respectively, $n$ is the number of data points per pulsation cycle, $\nu$ is the pulsation mode frequency and $\nu_{\rm samp}$ is the instrument's sampling frequency \citep{Murphy_PhD, Bowman_BOOK}. 
	
	A second correction arising from the difference in instrument passband is also needed, since the WASP passband is slightly more sensitive to bluer wavelengths compared to {\it Kepler}, such that observed pulsation amplitudes of A and F stars are (slightly) larger when observed at bluer wavelengths. We use the resultant value for this passband correction from \citet{Bowman2015a}, who estimated this correction factor to be 7.6~per~cent for the late-A star KIC~7106205 --- i.e. pulsation mode amplitudes are 7.6~per~cent smaller in \Kepler observations compared to WASP observations because of the difference in effective wavelength range of their passbands\footnote{More generally, this factor depends on the effective temperature of the star and the pulsation mode geometry, but for A and F stars the difference between \Kepler and WASP amplitudes is of order 10~per~cent.}. 
	
	Thus we corrected the WASP and truncated \Kepler data obtained in 2009 and 2010 for their respective instrument's integration time, and then transformed the WASP pulsation mode amplitudes into the \Kepler passband using the passband correction factor from \citet{Bowman2015a}. The remaining difference between the pulsation mode amplitudes observed by WASP and \Kepler in 2009 and 2010 is because of flux dilution from nearby and background stars within the larger WASP aperture \citep{Bowman2015a}. A ratio of the respectively corrected WASP and \Kepler mode amplitudes in 2009 and 2010 reveals an average dilution factor of $2.2 \pm 0.1$ --- i.e. the integration time and passband corrected WASP amplitudes are approximately 45~per~cent those of the \Kepler amplitudes. Of the integration time, passband and dilution correction factors, the latter is certainly the largest effect and the largest source of uncertainty in this analysis. This is already evident in the non-corrected amplitude spectra shown in Fig.~\ref{figure: grid}.
	
	\begin{figure*}
	\centering
	\includegraphics[width=0.49\textwidth]{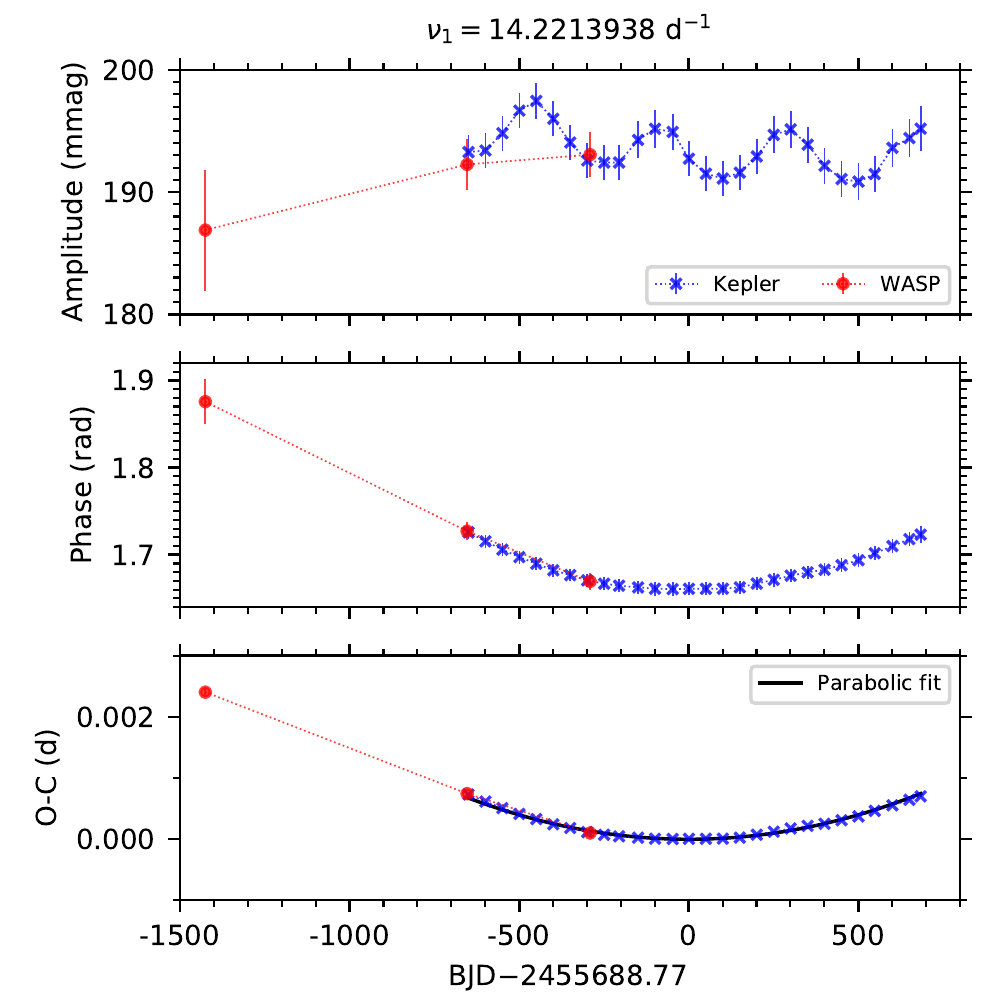}
	\includegraphics[width=0.49\textwidth]{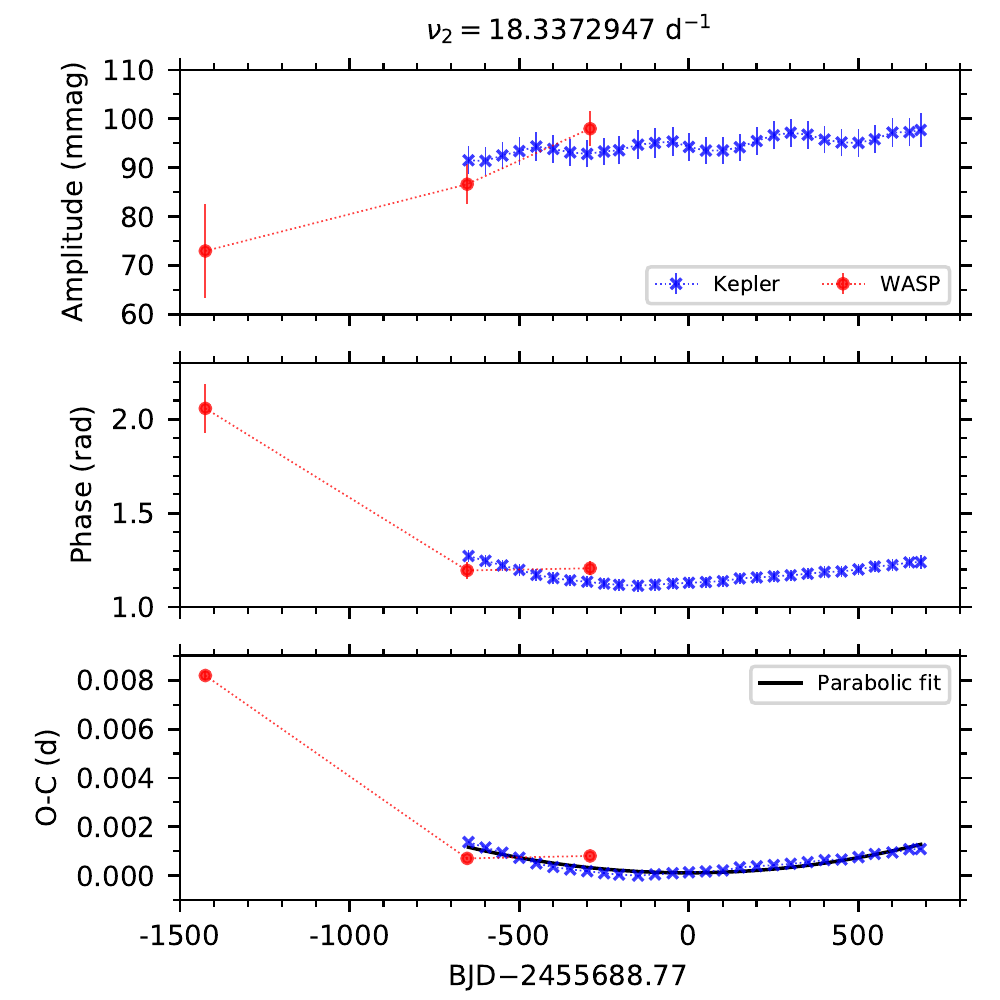}
	\caption{Inclusion of WASP data with \Kepler data to study the long-term amplitude and phase modulation of the fundamental, $\nu_1$ (left), and first overtone, $\nu_2$ (right), radial modes in KIC~5950759. Top panels: amplitude modulation in KIC~5950759 using the available LC \Kepler data (in bins of 150~d and 100-d overlap) as blue crosses and the three WASP seasons as red circles, which have been corrected for their respective integration times, passbands and dilution (see text for details). Middle panels: phase modulation tracked at fixed frequency. Bottom panels: O-C diagram in which the parabolic fit reveals the linear rate of change in period for each pulsation mode. In all panels, the zero-point of the time scale has been taken as the centre of the LC \Kepler data set (i.e. BJD~2455688.770). Note that the bottom two rows are equivalent.}
	\label{figure: WASP}
	\end{figure*}
	
	After the integration time, passband and dilution correction factors were determined and applied to the 2009 and 2010 WASP data, they were also applied to the 2007 WASP data to estimate the amplitude of the fundamental and first overtone radial modes in KIC~5950759. We provide the results for the corrected WASP amplitudes of $\nu_1$ and $\nu_2$ in the top-left and top-right panels of Fig.~\ref{figure: WASP}, respectively. We also divided the 4-yr \Kepler data into bins of 150~d with 100-d overlap, following the technique employed by \citet{Bowman2016a}, and show the amplitude and phase as a function of fixed frequency (cf. Table~\ref{table: freqs}) in Fig.~\ref{figure: WASP}. The quasi-sinusoidal variability in the pulsation amplitudes in the LC \Kepler data has a period equal to that of the \Kepler orbital period, and is caused by a changing signal-to-background flux ratio caused by the rotation of the spacecraft every 90~d. Thus, it is of instrumental and not astrophysical origin, although the amplitude is arguably constant within 2$\sigma$ regardless. Despite the relatively large amplitude uncertainties associated with the WASP data in 2007, we conclude that there is tentative evidence for a growing amplitude in the fundamental and first overtone radial modes in KIC~5950759 prior to the start of the \Kepler observations.

%	%	%	%	%	%	%	%	%	%
	
	\subsection{Period Changes}
	\label{subsection: period changes}
	
	During the main sequence and post-main sequence phases of stellar evolution, an intermediate mass star experiences an increase in radius, which increases the size of the acoustic cavity of its radial pulsation modes, hence the periods of radial modes increase with age. On the other hand, the relatively short-lived overall contraction phase near the TAMS, which results in a ``blue hook'' in the HR diagram for intermediate mass stars \citep{Iben1967a, Kippenhahn_BOOK}, is a phase of evolution in which the stellar radius decreases such that the periods of radial modes are expected to decrease. Consequently, the rate of change in the periods of radial modes theoretically allow stellar evolution to be measured \citep{Walraven1992, Breger1998c}. 
	
	Period changes in pulsation modes have been historically studied by means of the $\rm{O}-\rm{C}$ technique \citep{Percy1980e, Sterken2005a}. Based on the work of \citet{Breger1998c}, period changes in pulsation modes can be determined from
	
	\begin{equation}
	{\rm O} - {\rm C} = \frac{1}{2} \left( \frac{1}{P} ~ \frac{{\rm d}P}{{\rm d}t} \right) t^{2} = \frac{1}{2} \left( \frac{\dot{P}}{P} \right) t^{2} ~ ,
	\label{equation: period changes}
	\end{equation}
	
	\noindent where $\rm{O} - \rm{C}$ is the difference between the observed and calculated times of maxima in days, $t$ is the time span of the observations in days, and $P$ is the period of the signal in days, such that to convert the period change into its conventional units of yr$^{-1}$ one uses the number of days in a year (i.e. 365.2422). Thus, Eq.~(\ref{equation: period changes}) dictates that a linearly changing period produces a quadratic change in $\rm{O} - \rm{C}$ \citep{Percy1980e, Breger1998c, Sterken2005a}. 
	
	In addition to studying the amplitude modulation of pulsation modes in KIC~5950759, the concurrent WASP and \Kepler observations in 2009 and 2010 allow us the unique opportunity to probe its phase (i.e. frequency) modulation. Using the technique employed by \citet{Bowman2016a}, we calculate the pulsation phase using a linear least-squares fit to each of the 150-d bins of LC \Kepler data, which is shown in the middle row of Fig.~\ref{figure: WASP}. Furthermore, to include the 2007 WASP data, as before we use the truncated 2009 and 2010 overlapping \Kepler data to compare the pulsation phase during the concurrent 2009 and 2010 WASP seasons. Since we are assuming and fixing the pre-optimised pulsation mode frequencies of $\nu_1$ and $\nu_2$ (cf. Table~\ref{table: freqs}), we perform an independent linear least-squares fit to each of the WASP and truncated \Kepler data in 2009 and 2010. We compare the corresponding pulsation phases and find a constant phase offset of $0.54 \pm 0.01$~rad between the 2009 WASP and truncated 2009 \Kepler data (and the same value between the 2010 WASP and truncated 2010 \Kepler data). Such a phase offset is understandable given the difference in passband between the two instruments and because pulsation phases depend on wavelength\footnote{Pulsation mode phase differences is a method of mode identification for multi-colour time series photometric observations --- see \citet{Watson1988} and \citet{Heynderickx1994b} for early applications of this technique.}.
	
	After correcting the 2007, 2009 and 2010 WASP phases for the passband phase offset value, we include them alongside the LC \Kepler data of KIC~5950759 in the middle panels of Fig.~\ref{figure: WASP}. A clear parabolic change in pulsation phase is evident for both the fundamental ($\nu_1$; middle-left panel of Fig.~\ref{figure: WASP}) and first overtone ($\nu_2$; middle-right panel of Fig.~\ref{figure: WASP}), which indicate a linearly increasing pulsation period. Under the reasonable assumption of negligible mass loss across a period of approximately 6~yr, this implies an increasing stellar radius in KIC~5950759. The observed phase modulation for each radial mode, $\nu_1$ and $\nu_2$, is converted using ${\rm O} - {\rm C} = \frac{\phi}{2\pi\nu}$, with the resultant ${\rm O}-{\rm C}$ diagrams shown in the bottom-left and bottom-right panels of Fig.~\ref{figure: WASP}, respectively. To avoid adding unnecessary additional uncertainty, we fit only the \Kepler data with Eq.~(\ref{equation: period changes}) with $\left( \frac{1}{P} ~ \frac{{\rm d}P}{{\rm d}t} \right)$ as a free parameter. The fitted fractional rate of change in period values have units of d$^{-1}$ since $t$ and ${\rm O}-{\rm C}$ are in units of d, we converted them into the conventional unit of yr$^{-1}$. This yielded best-fitting values of $(1.17 \pm 0.02) \times 10^{-6}$~yr$^{-1}$ and $(1.82 \pm 0.12) \times 10^{-6}$~yr$^{-1}$ for the fractional period changes of the fundamental and first overtone radial modes, respectively.
	
	Contrary to stellar evolution theory, approximately an equal fraction of Population~{\sc I} \dsct stars are observed to have increasing periods compared to those with decreasing periods in their radial modes \citep{Breger1990a, Guzik1991a}. The typical linear period changes observed in \dsct stars can be as large as ${\dot{P}}/{P} \simeq 10^{-7}$~yr$^{-1}$, which is at least an order of magnitude larger than those predicted by stellar evolution theory for main sequence \dsct stars \citep{Breger1998c}. Similarly large period changes have also been observed in HADS stars (e.g. \citealt{Walraven1992}) and SX~Phe stars (e.g. \citealt{Murphy2013b}). In KIC~5950759, we measure even larger period changes of order 10$^{-6}$~yr$^{-1}$, which could be used to infer that this HADS star is in a relatively short-lived stage of stellar evolution, such as near or beyond the TAMS, if such variability is caused by evolution. The increasing period of the radial modes also implies that the radius of KIC~5950759 is increasing, which suggests a post-main sequence stage of evolution as opposed to the overall contraction phase. An order of magnitude estimate of the change in radius using the period-density relation for classical pulsators (see e.g. \citealt{Stellingwerf1979}) and assuming a constant mass of $M=1.6$~M$_{\odot}$ yields $\dot{R} \simeq 1$~km\,yr$^{-1}$, but the uncertainties on this are substantial. Such a value is three orders of magnitude larger than expected from stellar evolution models. Hence it is unlikely that the observed period (and implied radius) changes are caused by evolution.
	
	We note that the amplitude and phase modulation observed in KIC~5950759 are seemingly unique, because phase modulation in such stars appears as jumps rather than smooth variation over time (see e.g. \citealt{Breger1998c, Boonyarak2011a}). On the other hand, few HADS stars have been studied using such high quality photometric time series as those provided by the \Kepler mission. Since the observed phase modulation is much larger than predicted by evolution models, this indicates that the short-term evolution of some stars can be quite {\it `noisy'}, such that period changes measured over time spans of years and decades may be caused by physical processes that are not resolved by the large time steps taken in evolutionary calculations.

%%%%%%%%%%%%%%%%%%%%%%%%%%%%%%%%%%%%%%%%%%%%%%%%%%

\section{Spectroscopy}
\label{section: spectroscopy}

\begin{figure*}
\centering
\includegraphics[width=0.99\textwidth]{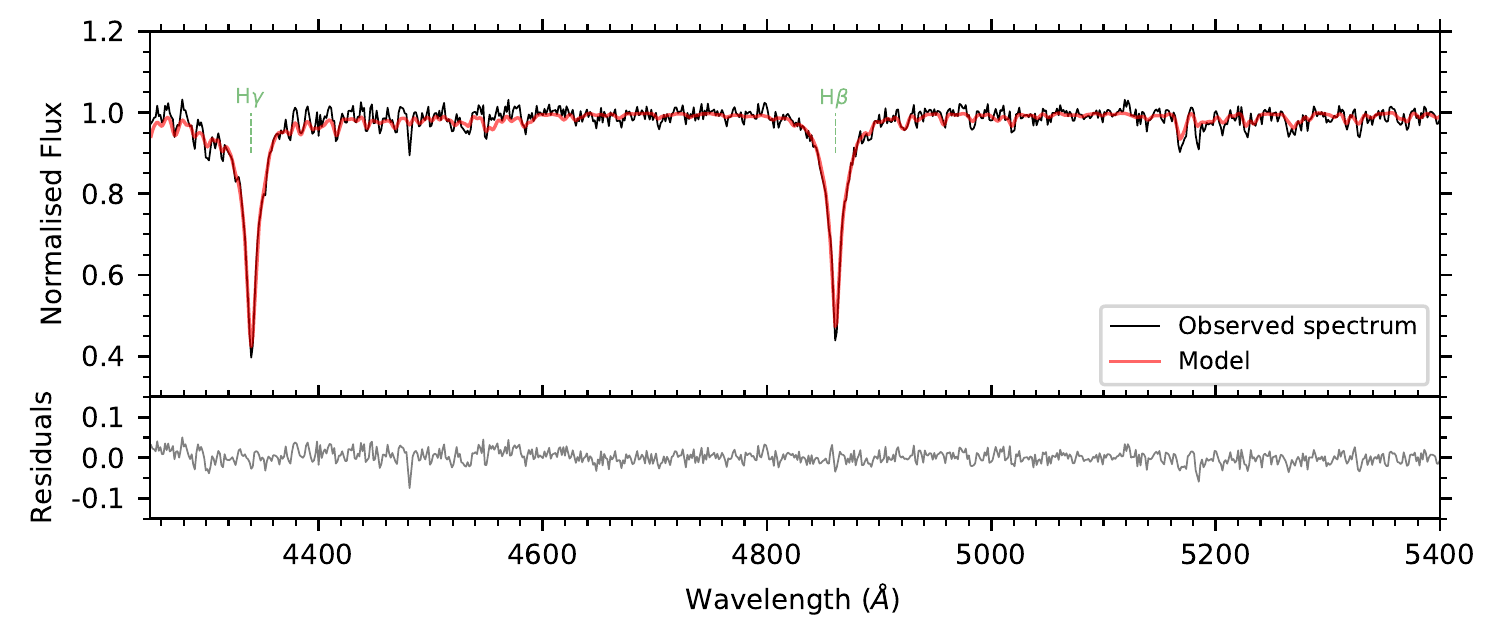}
\caption{Top panel: observed spectrum of KIC~5950759 (black) and the best fitting model (without continuum correction) obtained using the {\sc GSSP} software \citep{Tkachenko2015b} in red. Bottom panel: residuals of model and observations.}
\label{figure: spectrum}
\end{figure*}

KIC~5950759 is a relatively faint star within the nominal \Kepler field of view with $V = 13.516$~mag \citep{Pigulski2009}. To determine atmospheric parameters, and in particular an accurate effective temperature, $T_{\rm eff}$, we obtained a spectrum on 2016-09-28 using the Intermediate Dispersion Spectrograph (IDS) mounted on the 2.5-m Isaac Newton Telescope (INT) located at the Roque de los Muchachos Observatory on the island of La Palma, Spain. The spectrum was obtained using the R400V grating with a resolving power of $R \simeq 1500$ at 4500~\AA. We reduced the data (including bias subtraction, flat fielding and wavelength calibration) using {\sc starlink}\footnote{\url{http://starlink.eao.hawaii.edu/starlink/}} software \citep{STARLINK_2014}. We normalised the spectrum by fitting a spline through manually selected continuum points. A section of the normalised reduced IDS spectrum of KIC~5950759 is shown in Fig.~\ref{figure: spectrum}. Spectral classification indicates that KIC~5950759 is a slow rotator from its narrow spectral lines, and is also a metal-poor A-type star given the lack of strong metal lines \citep{GRAY_BOOK, GRAY_CORBALLY_BOOK}.

To analyse the spectrum of KIC~5950759 using a quantitative method and extract atmospheric parameters, we used the Grid Search in Stellar Parameters ({\sc GSSP}) software \citep{Tkachenko2015b}. The {\sc GSSP} code computes a grid of synthetic spectra across a range of values in effective temperature ($T_{\rm eff}$), surface gravity ($\log\,g$), metallicity ([M/H]), and projected surface rotational velocity ($v\,\sin\,i$). The grid of synthetic spectra is computed using the {\sc SynthV} LTE radiative transfer code \citep{Tsymbal1996} and atmospheric models computed from the {\sc LLmodels} code \citep{Shulyak2004}. In our analysis we fix the microturbutent and macroturbutent velocities to 2.0 and 0.0~km\,s$^{-1}$, respectively, which are reasonable for stars of spectral type A \citep{GRAY_BOOK, Tkachenko2013a, Niemczura2015}. As mentioned previously, the low resolution of the observed spectrum makes it difficult to constrain the surface gravity and rotation rate of KIC~5950759. Furthermore, for stars below approximately 8000~K, the wings of Balmer lines are not sensitive to the surface gravity, which means that metal lines are typically used instead \citep{GRAY_BOOK, Smalley2005a}. Given the resolving power of the IDS/INT instrument and the resultant resolution of the observed spectrum, rotational and instrumental broadening are indistinguishable for $v\,\sin\,i$ values below some 50~km\,s$^{-1}$.

We shifted the observed spectrum to rest wavelengths and used the {\sc GSSP} code to quantitatively compare it to each spectrum within the grid of synthetic spectra by means of minimising the corresponding $\chi^2$ function. After some initial tests, we confirmed that KIC~5950759 is a slow rotator, with $v\sin\,i \leq 50$~km\,s$^{-1}$, such that the inferred rotational broadening was smaller than the instrumental broadening. Therefore, we set the velocity broadening component to 0~km\,s$^{-1}$ when using {\sc GSSP} in our subsequent analysis. As a first step we use the 4250--5400\,{\AA} wavelength range to constrain the effective temperature to $T_{\rm eff} = 7340 \pm 230$~K, which is primarily driven by the shape of the $H\gamma$ and $H\beta$ Balmer lines in this range. Given the low resolution of the spectrum and lack of strong metal lines, we are unable to constrain the surface gravity to better than $\log\,g = 2.9 \pm 1.3$~dex and metallicity to [M/H]~$= -1.18 \pm 0.30$~dex. As a second step, we performed a second fitting scheme including a global continuum correction factor, which is assumed to be wavelength-independent and is computed in {\sc GSSP} from the least-squares fit of the synthetic to the observed spectrum of the star. For the same wavelength regime of 4250-5400\,{\AA}, we constrained the effective temperature to $T_{\rm eff} = 7470 \pm 255$~K, the surface gravity to $\log\,g = 3.6 \pm 1.2$~dex and the metallicity to [M/H]~$= -0.65 \pm 0.45$~dex. These parameters are consistent with those from our previous analysis, and are provided in Table~\ref{table: parameters}. 

As an additional confirmation of the sub-solar metallicity of KIC~5950759, we estimated the metallicity using the methodology of \citet{Smalley1993b}. An estimate of the line blocking in the 4600-4700~{\AA} region using a model with $T_{\rm eff} = 7500$~K and $\log\,g = 3.5$ yielded a metallicity of [M/H]~$= -1.1 \pm 0.4$, which is consistent with the sub-solar metallicity values obtained from {\sc GSSP}. The high frequency of the fundamental radial mode in KIC~5950759, and the relatively large observed frequency ratio of $0.77554$ also support that KIC~5950759 is a metal-poor star (see e.g. \citealt{Petersen1996, Poretti2003b, Murphy2013b, Daszy2020c}).

%%%%%%%%%%%%%%%%%%%%%%%%%%%%%%%%%%%%%%%%%%%%%%%%%%

\section{Estimating the luminosity from GAIA}
\label{section: gaia}
	
In their study of the period-luminosity (P-L) relation for \dsct stars observed by \Kepler with available Gaia parallaxes, \citet{Ziaali2019a} include KIC~5950759 as an example of a star known to pulsate in the fundamental radial mode (see their figure~3). Interestingly, KIC~5950759 does not fall on the derived P-L relationship. Owing to the pulsation period being known to an extremely high precision, the discrepancy may be caused by an incorrect luminosity from an incorrect estimate of the interstellar reddening and/or distance. \citet{Murphy2019a} calculated luminosities for a large sample of more than 2000 \dsct stars observed by the \Kepler mission with available Gaia DR2 parallaxes \citep{Gaia2016a, Gaia2018a}, distances from \citet{Bailer-Jones2018c} and dust maps from \citet{Green_G_2018a}. From their ensemble analysis, \citet{Murphy2019a} calculated a luminosity of $\log\,(L/{\rm L_{\odot}}) = 1.017 \pm 0.038$ for KIC~5950759 assuming an effective temperature $T_{\rm eff} = 8070 \pm 280$~K.

The bolometric corrections of \citet{Murphy2019a} were calculated based on parameters available from the \citet{Mathur2017a} stellar properties catalogue, which are somewhat consistent with those found in this work except for the metallicity, which we find to be sub-solar with $-1.18 <$~[M/H]~$< -0.65$~dex (cf Table~\ref{table: parameters}). This can have a large effect on the bolometric correction, so we have re-calculated the luminosity using the updated astrometry from Gaia EDR3 \citep{Gaia2020a*}, the effective temperatures, surface gravities and metallicities from our spectroscopic analysis, which yields $\log\,(L/{\rm L}_{\odot}) = 1.064 \pm 0.045$. The fractional distance uncertainty for KIC~5950759 is $\sim$4~per~cent, whereas the fractional uncertainty on the 0.22 mag extinction value is 16~per~cent. Thus neither of these parameters is a likely source of the $\simeq$~0.2~mag difference between the calculated absolute magnitude of KIC~5950759 and the inferred P-L relation in \citet{Ziaali2019a}.

A potential cause of a star appearing too bright in the P-L diagram is binarity. The Gaia renormalised unit weight error (RUWE) has been interpreted as an indication of binarity, with RUWE~$>2$ used to suggest binarity \citep{Evans_D_2018, Rizzuto2018}. In Gaia EDR3, the RUWE parameter for KIC~5950759 is 1.047, so it is probably not a binary. Furthermore, after comparing the Gaia DR2 astrometric excess noise of 0.11\,mas for KIC~5950759 to the average DR2 excess noise of 0.14\,mas for the large sample of \Kepler A and F stars from \citet{Murphy2019a}, we conclude that KIC~5950759 is unlikely to be a binary.

More importantly, \citet{McNamara2011b} found that corrections to the P-L relation were required for low-metallicity radial-mode pulsators. For stars of [Fe/H]~$\simeq -1.0$, a correction of 0.2~mag is required, such that a star is over-luminous by this amount compared to the luminosity predicted solely based on its dominant pulsation period \citep{McNamara2011b}. This correction would account for the location of KIC~5950759 being above the P-L relation found by \citet{Ziaali2019a} for \Kepler \dsct stars. On the other hand, we cannot exclude an incorrect extinction for KIC~5950759 given its faint apparent brightness and larger inferred Gaia distance compared to typical A and F stars in the \Kepler field.

Finally, we note that the large pulsation amplitude of KIC~5950759 causes substantial temperature, radius, and radial velocity variations during the pulsation cycle. Using equation~5 of \citet{Kjeldsen1995}, we estimate radial velocity variability of $\sim$100~km\,s$^{-1}$ for measured brightness variations of $\pm$0.4\,mag, given that in \Kepler SC data the peak-to-peak light amplitude variations exceed 0.7~mag. We did not include the radius and temperature variations in the 10\,000-iteration Monte Carlo simulation for our improved calculation of the luminosity of KIC~5950759 in this work. Yet, we did include a 1$\sigma$ apparent magnitude change of 0.14~mag calculated from the standard deviation of the \Kepler brightness measurements.

%%%%%%%%%%%%%%%%%%%%%%%%%%%%%%%%%%%%%%%%%%%%%%%%%%

\section{Asteroseismic Modelling}
\label{section: modelling}

To ascertain the mass and evolutionary stage of KIC~5950759, we performed forward seismic modelling based on its observed pulsation mode frequencies, spectroscopic parameters and Gaia luminosity. Since \Kepler data are of exceptionally high quality, the two dominant pulsation mode frequencies and their frequency ratio are known to a very high precision. The high amplitudes of the pulsation modes and their frequency ratio indicate that they are likely the fundamental and first overtone radial modes, so a sensible method to model KIC~5950759 is by using Petersen diagrams. The frequency and frequency ratio of radial modes depend primarily on the mass, age, evolutionary stage and metallicity. Thus, Petersen diagrams offer a useful diagnostic in constraining these parameters of radial pulsators (see e.g. \citealt{Petersen1973, Petersen1996, Daszy2020c}).

We calculated evolution models with the {\sc Warsaw-NewJersey} code (see \citealt{Pamyat1998c, Pamyat1999b}) using the {\sc OPAL} opacity tables \citep{Iglesias1996} and a solar chemical mixture \citep{Asplund2009} for various metallicity values. Rotation, or more specifically the mean effect of the centrifugal force, is included within the {\sc Warsaw-NewJersey} code, and solid-body rotation and constant global angular momentum during evolution are assumed. Convection in the stellar envelope is based on standard mixing length theory, and the convective flux is assumed to be constant throughout a pulsation cycle. This is more commonly known as the convective flux freezing approximation, which is robust if convection is not efficient, such as in the envelopes of early type stars. Linear non-adiabatic pulsation models were calculated using the code of \citet{Dziembowski1977b}.

\begin{figure}
\centering
\includegraphics[width=0.99\columnwidth]{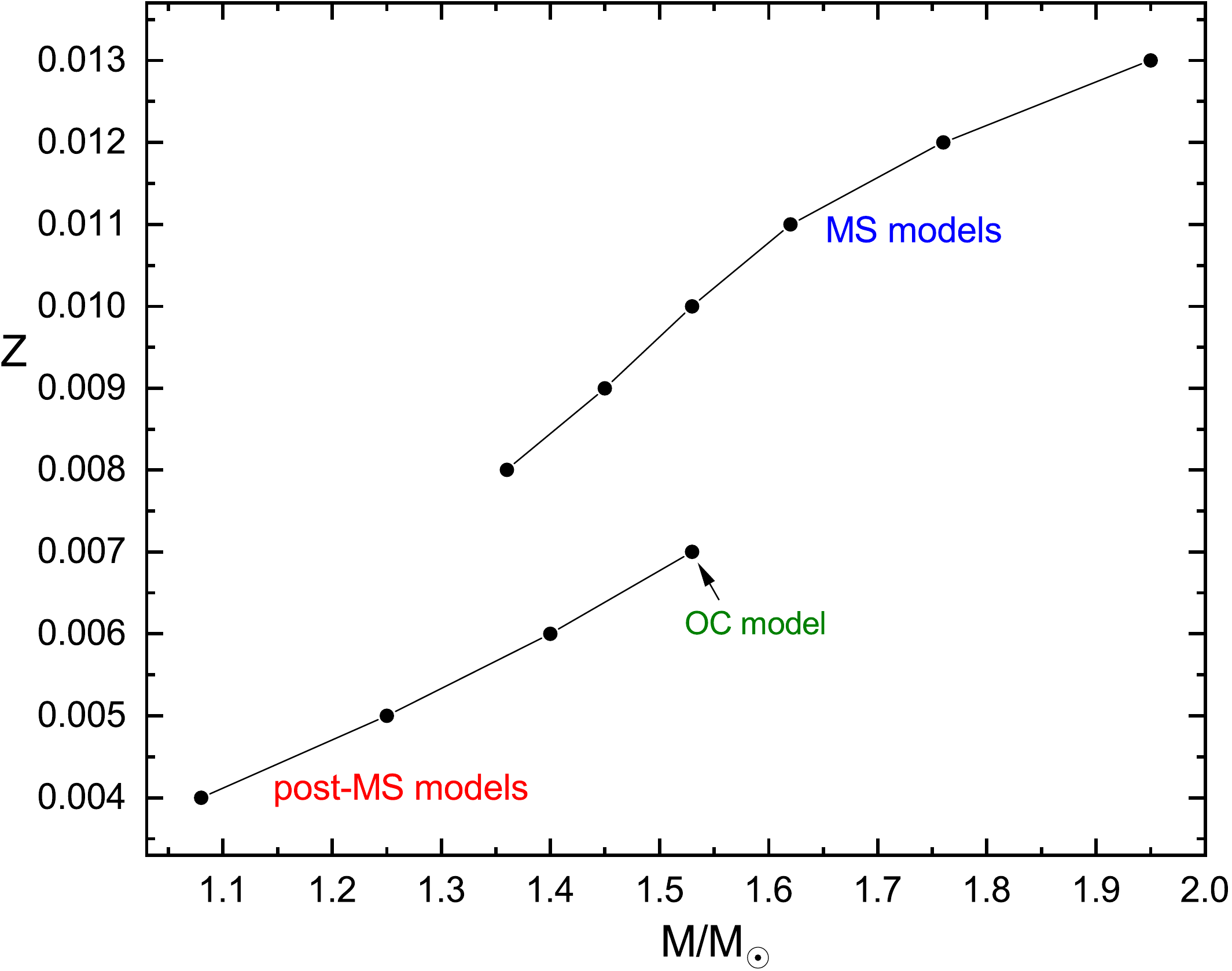}
\caption{Models calculated across a range in metallicity which match the observed frequency ratio observed in KIC~5950759. The strong constraint of the observed frequency ratio creates two branches in terms of possible evolutionary phases (where MS = main sequence, OC = overall contraction phase, post-MS = post main sequence).}
\label{figure: M-Z plane}
\end{figure}

\begin{figure}
\centering
\includegraphics[width=0.97\columnwidth]{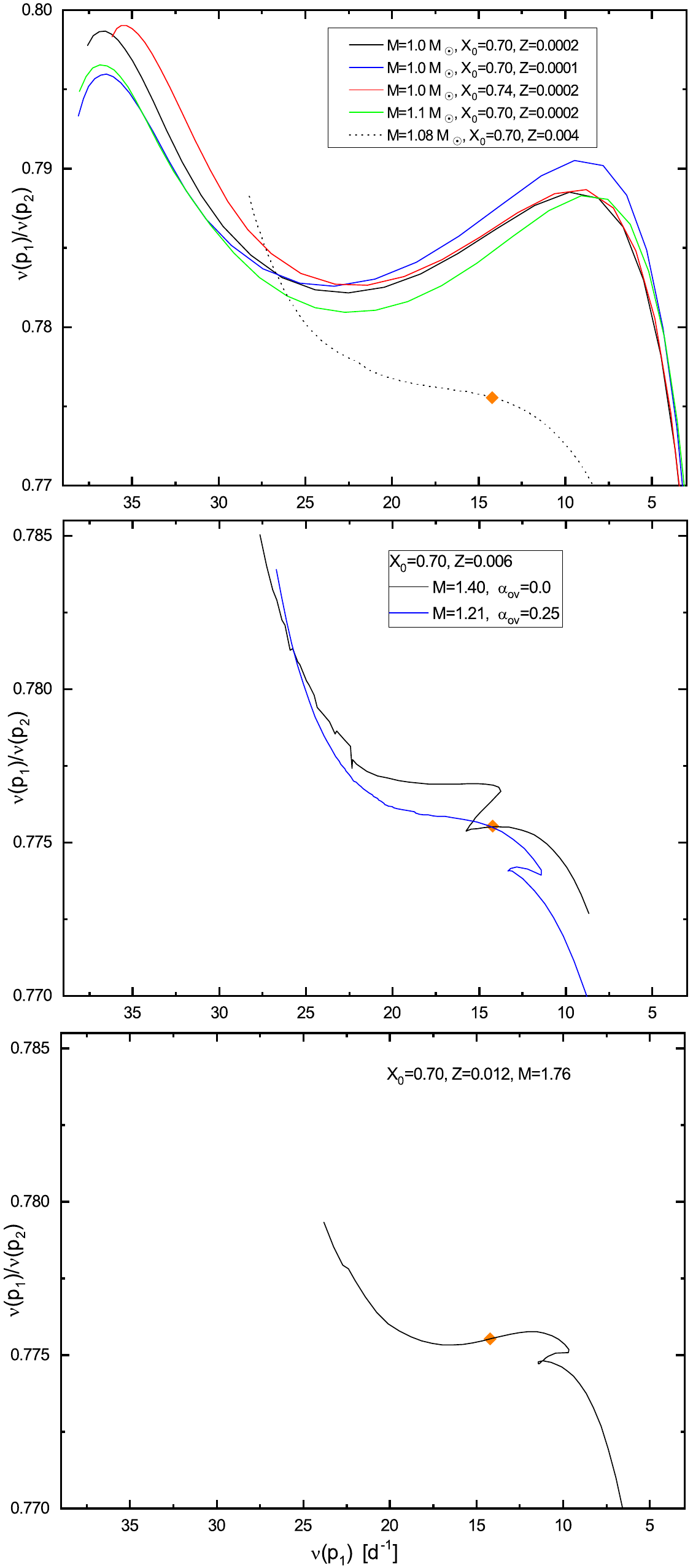}
\caption{Petersen diagrams spanning from the zero-age main sequence to the immediate post-main sequence phase of evolution for various metallicity values in our grid. In each panel, the location of the best model is indicated by the orange diamond. The observed frequency ratio of $\nu_1/\nu_2$ for KIC~5950759 clearly excludes low metallicities (i.e. $Z < 0.003$).}
\label{figure: Petersen}
\end{figure}

As described in section~\ref{section: spectroscopy}, there are large uncertainties associated with determining the metallicity of a slowly rotating, high-amplitude pulsating star with a low resolution spectrum. Therefore we chose to survey a range of models with different metallicities between $0.004 \leq Z \leq 0.013$~dex in steps of 0.001~dex, which is larger than the inferred 1$\sigma$ metallicity uncertainty from spectroscopy, and determine the best fitting mass and age. For each value of metallicity, we found the model that fits the dominant frequency ($\nu_1$) identified as the fundamental radial mode, and closely matches the frequency identified as the first overtone radial mode ($\nu_2$). Our results are shown in the mass-metallicity plane in Fig.~\ref{figure: M-Z plane}. All the models in Fig.~\ref{figure: M-Z plane} match the observed frequencies $\nu_1$ and $\nu_2$, which demonstrates the strong mass-metallicity degeneracy. In general, fitting the two pulsation modes alone is unable to break the degeneracy between mass and age\footnote{Analogous to isochrones (lines of constant age), one can fit isopycnals or isopycnics (i.e. lines or surfaces of constant density) using frequency ratios of radial modes (see \citealt{Hermans_MSc}) to serve as a proxy for stellar age in the HR~diagram as opposed to using uncertain spectroscopic $\log\,g$ values.}. Hence one needs additional information, such as fitting a third mode or including spectroscopic constraints, to constrain the age of a star. Specifically, effective temperature and metallicity derived from spectroscopy and the luminosity estimated from Gaia are used to delimit the possible parameter space in the HR~diagram.

% - - - - - % - - - - - % - - - - - % - - - - - %
\begin{table*}
\centering
\caption{The parameters of the models that reproduce exactly the observed dominant frequency $\nu_1$ as the radial fundamental mode ($\ell=0,~p_1$) and closely match the observed frequency $\nu_2$ as the radial first overtone ($\ell=0,~p_2$). All models have an initial hydrogen abundance of $X_{0}=0.70$, mixing length parameter $\alpha_{\rm MLT}=0.1$ and an initial rotational velocity $v_{\rm rot}=10~\kms$. Columns are metallicity ($Z$), the convective core overshooting ($\alpha_{\rm ov}$), mass ($M$), phase of evolution (where OC denotes the overall contraction phase), age, effective temperature ($\log T_{\rm eff}$), luminosity ($\log L/L_{\odot}$), the theoretical frequency of the radial first overtone $\nu(p_2)$, the frequency ratio of the first two radial modes and their normalised instability parameters $\eta(p_1)$ and $\eta(p_2)$, and the theoretically predicted fractional period change of the fundamental radial mode.}
\begin{tabular}{cccccccccccc}
\hline
$Z$	&	$\alpha_{\rm ov}$	&	$M$	&	phase	&	age	&	$\log T_{\rm eff}$	&	$\log L/{\rm L}_{\odot}$	&	$\nu(p_2)$	&	$\nu(p_1)/\nu(p_2)$	&	$\eta(p_1)$	&	$\eta(p_2)$	&	${\dot{P}}/{P}$	\\
	&	&	(M$_{\odot}$)	&	&	(Gyr)	&	&	&	(d$^{-1}$)	&	&	&	&	(yr$^{-1}$)	\\
\hline
0.004	&	0.0	&	1.08	&	post-MS	&	3.66	&	3.8285	&	0.743	&	18.33686	&	0.77556	&	$+0.079$	& 	$+0.036$		&	$+8.5 \times 10^{-10}$	\\
0.005	&	0.0	&	1.25	&	post-MS	&	2.25	&	3.8652	&	0.932	&	18.33704	&	0.77556	&	$+0.068$	&	$+0.066$		&	$+1.2 \times 10^{-9}$	\\
0.006	&	0.0	&	1.40	&	post-MS	&	1.62	&	3.8919	&	1.071	&	18.33817	&	0.77551	&	$-0.029$	&	$+0.015$		&	$+1.0 \times 10^{-9}$	\\
0.006	&	0.25	&	1.21	&	MS		&	2.60	&	3.8195	&	0.737	&	18.33857	&	0.77549	&	$+0.070$	&	$+0.024$		&	$+6.1 \times 10^{-10}$	\\
0.007	&	0.0	&	1.53	&	OC		&	1.31	&	3.9099	&	1.167	&	18.33867	&	0.77549	&	$-0.143$	&	$-0.071$		&	$-2.3 \times 10^{-8}$		\\
0.008	&	0.0	&	1.36	&	MS		&	1.73	&	3.8359	&	0.835	&	18.33685	&	0.77556	&	$+0.082$	&	$+0.045$		&	$+7.7 \times 10^{-10}$		\\
0.009	&	0.0	&	1.45	&	MS		&	1.42	&	3.8510	&	0.914	&	18.33668	&	0.77557	&	$+0.079$	&	$+0.058$		&	$+8.2 \times 10^{-10}$		\\
0.010	&	0.0	&	1.53	&	MS		&	1.21	&	3.8637	&	0.979	&	18.33759	&	0.77553	&	$+0.059$	&	$+0.058$		&	$+8.8 \times 10^{-10}$		\\
0.011	&	0.0	&	1.62	&	MS		&	1.02	&	3.8775	&	1.051	&	18.33797	&	0.77552	&	$+0.016$	&	$+0.038$		&	$+9.4 \times 10^{-10}$		\\
0.012	&	0.0	&	1.76	&	MS		&	0.78	&	3.9014	&	1.170	&	18.33771	&	0.77555	&	$-0.104$	&	$-0.048$		&	$+1.1 \times 10^{-9}$	\\
0.013	&	0.0	&	1.95	&	MS		&	0.56	&	3.9324	&	1.323	&	18.33985	&	0.77544	&	$-0.326$	&	$-0.256$		&	$+1.3 \times 10^{-9}$	\\
\hline
\end{tabular}
\label{table: models}
\end{table*}
% - - - - - % - - - - - % - - - - - % - - - - - %

As can be seen in Fig.~\ref{figure: M-Z plane}, there are two branches of possible solutions for KIC~5950759: main-sequence models for which $Z \ge 0.008$ and post-main sequence models for which $0.004 \leq Z < 0.007$. For each model at a given metallicity shown in Fig.~\ref{figure: M-Z plane}, we provide the best-fitting mass, age, effective temperature, luminosity, theoretical frequency of the first overtone radial mode, and the frequency ratio of the fundamental and first overtone radial modes in Table~\ref{table: models}. For each best-fitting model, we also calculate the normalised instability parameters $\eta(p_1)$ and $\eta(p_2)$ (as defined by \citealt{Stellingwerf1978c}; see also \citealt{Daszy2020c}), for which positive values indicate that a pulsation mode is excited (i.e. unstable) and negative values indicate it is not (i.e. stable).

\begin{figure*}
\centering
\includegraphics[width=0.99\textwidth]{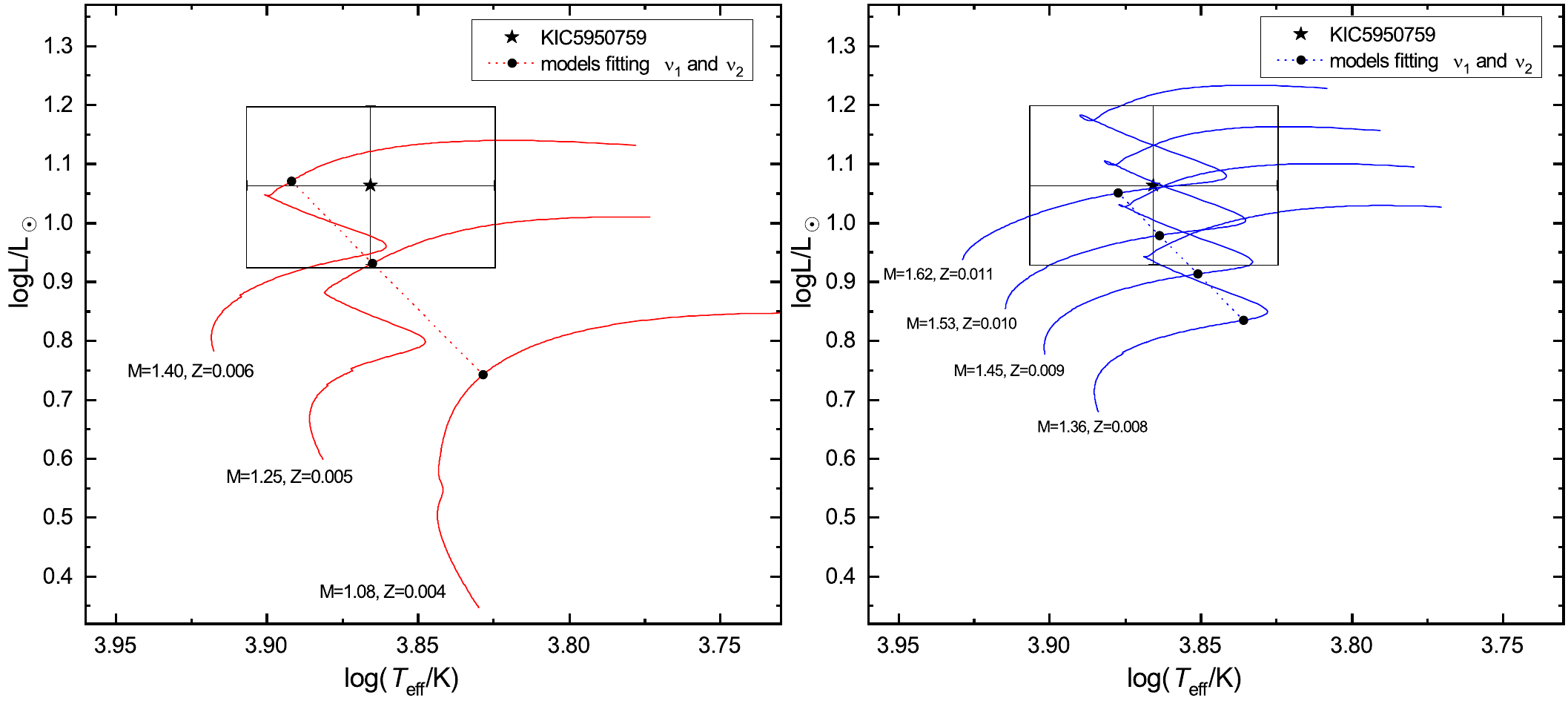}
\caption{HR~diagram with the 3$\sigma$ $T_{\rm eff}-\log\,L$ error box of KIC~5950759 and evolutionary tracks computed for various masses and metallicities at the adopted values of an initial hydrogen abundance $X_{0} = 0.7$, the mixing length parameter $\alpha_{\rm MLT} = 0.1$ and no convective core overshooting. Seismic models that best reproduce the two dominant frequencies as the fundamental and first overtone radial modes are marked with dots. In the left panel, the best-fitting seismic models are in the post-main sequence phase of evolution and in the right panel they are in main sequence phase of evolution.}
\label{figure: HRD}
\end{figure*}

However, not all models shown in Fig.~\ref{figure: M-Z plane} lie within 3$\sigma$ of the spectroscopic constraints on the effective temperature and Gaia luminosity for KIC~5950759. To demonstrate the importance of these additional constraints in forward seismic modelling, we show Petersen diagrams for different parameter combinations in Fig.~\ref{figure: Petersen}. In the top panel, Petersen diagrams for extremely metal-poor stars are shown as coloured lines; none of which match the observed frequency ratio of KIC~5950759. On the contrary, the observed frequency ratio excludes the possibility of KIC~5950759 having $Z < 0.003$, with the lowest value in our grid that matches the frequency ratio being $Z = 0.004$. At the other end of the metallicity scale are main sequence models with higher masses and younger ages that match the observed frequency ratio of KIC~5950759. We show an example of a near-solar metallicity (i.e. $Z=0.012$) Petersen diagram in the bottom panel of Fig.~\ref{figure: Petersen}. Therefore, the metallicity of KIC~5950759 must lie within the range of $0.004 \leq Z \leq 0.013$, if one only considers the fit using only the fundamental and first overtone radial mode frequencies. In Fig.~\ref{figure: HRD}, we show the position of KIC~5950759 with its 3$\sigma$ error box and seismic models with the corresponding evolutionary tracks for several metallicities. In the left panel we marked seismic models in the post-main sequence phase of evolution and in the right panel seismic models in main sequence phase of evolution. The location of KIC~5950759 in the HR~diagram combined with the best fitting frequency ratio of its fundamental and first overtone radial modes indicates a relatively low mass star ($M \leq 1.6$~M$_{\odot}$) with sub-solar metallicity near the overall contraction phase and the TAMS.

We also explored including convective core overshooting as a parameter in our models. The {\sc Warsaw-NewJersey} code uses a two parameter prescription introduced by \citet{Dziembowski2008}, which allows for a non-zero gradient of the hydrogen abundance inside the partly mixed region above the convective core. An example of the effect of overshooting with the efficiency parameter $\alpha_{\rm ov} = 0.25$ for $Z=0.006$ is shown in the middle panel of Fig.~\ref{figure: Petersen}. Increasing the amount of near-core mixing provides the convective core with additional hydrogen available for nuclear burning during the main sequence. This extends the main sequence lifetime of the star and moves the location of the TAMS and blue hook to larger radii and hence higher luminosities in the HR~diagram. Consequently, it also moves the TAMS to smaller frequency ratios in the Petersen diagram. Therefore, for $Z = 0.006$, as shown in the middle panel of Fig.~\ref{figure: Petersen} and listed in Table~\ref{table: models}, the best fitting model with $Z=0.006$ and $\alpha_{\rm ov}=0.00$ is a post-main sequence star, and with $Z=0.006$ and $\alpha_{\rm ov}=0.25$ it is a main sequence star.

A major advantage of using non-adiabatic pulsation models is that we are also able to determine if modes are excited. As can be seen in Table~\ref{table: models}, the normalised instability parameters, $\eta(p_1)$ and $\eta(p_2)$, are negative for $Z \geq 0.012$, which means we can exclude relatively high mass (i.e. $M \geq 1.7$~M$_{\odot}$) main sequence models as we do not expect the fundamental and first overtone radial modes to be excited. Furthermore, the best fitting model for $Z=0.007$ also has negative $\eta(p_1)$ and $\eta(p_2)$ values, which means that we can exclude the overall contraction phase. Combining the requirement for positive normalised instability parameters with the $3\sigma$ confidence intervals for $T_{\rm eff}$ and $\log\,L$ for KIC~5950759, the best fitting models are those with metallicity values approximately in the two ranges of $Z \in (0.005, 0.006)\cup(0.010, 0.011)$. However, our best-fitting models consistently place KIC~5950759 near to the TAMS\footnote{Note that the TAMS as defined by \citet{Iben1967a} is before the overall contraction phase and blue hook in the HR~diagram, such that $X_{c} \simeq 0.05$. Whereas other studies, such as \citet{Dotter2016a}, define the TAMS as when the core hydrogen content (numerically) reaches zero.}. Given this result and the limitations and uncertainties associated with the determination of $T_{\rm eff}$, $v\,\sin\,i$, and $Z$ for KIC~5950759, it is highly desirable that high-resolution spectra of KIC~5950759 be obtained. Preferably spectra sampling the pulsation cycle of $\nu_1$ should be obtained. This will allow a more accurate determination of the metallicity and effective temperature given that the large amplitude pulsations of KIC~5950759 are likely causing significant line broadening which cannot be taken into account in the analysis of only a single spectrum.

Motivated by the large observed values for the fractional period change in the fundamental and first overtone radial modes, we also checked the theoretically predicted values within our models for agreement. For completeness we extracted the fractional period change of the fundamental radial mode of the best-fitting model at each metallicity value in our grid. These values are provided in the last column of Table~\ref{table: models}. The values are similar for the first overtone radial mode, and also comparable to those calculated by \citet{Breger1998c}. Yet, all predicted values are considerably smaller than the fractional period change observed in KIC~5950759 by at least two orders of magnitude. It is well established that \dsct stars (both Population~{\sc I} and {\sc II}) have observed period changes much larger than predicted by evolutionary models \citep{Rod1995d, Rod2001, Breger1998c, Bowman2016a}. Therefore, we conclude that the (common) observed period changes in \dsct pulsators are not the result of stellar evolution, but are likely connected to the inherent non-linear excitation mechanism of high-amplitude pulsation modes, and the interactions of modes causing modulated amplitudes and frequencies over time scales of years and decades (e.g. resonant mode coupling; \citealt{Dziembowski1985a, Moskalik1985, Breger2014, Bowman2016a}). Therefore, our results demonstrate the need for future studies to move beyond linear asteroseismology, and include additional constraints based on pulsation amplitudes and non-linear effects at work in \dsct stars.

%%%%%%%%%%%%%%%%%%%%%%%%%%%%%%%%%%%%%%%%%%%%%%%%%%

\section{Conclusions}
\label{section: conclusions}

In this work we have performed photometric, spectroscopic and modelling analyses of the high-amplitude \dsct star KIC~5950759. Originally identified as having unique properties owing to its long term and large fractional period change in its fundamental radial and first overtone radial modes by \citet{Bowman_PhD}, our detailed frequency analysis of the \Kepler mission photometry of KIC~5950759 reveals additional non-radial pulsation modes. We extend the 4-yr time span of the \Kepler mission using WASP data, and demonstrate that KIC~5950759 exhibits long-term amplitude and phase modulation in its two dominant radial modes. The measured fractional linearly increasing periods of the fundamental and first overtone radial mode are of order $10^{-6}$~yr$^{-1}$, which are at least two orders of magnitude larger than predicted for evolutionary changes in main sequence and post-main sequence stars. Given such a large discrepancy is not supported by stellar evolution models, we conclude that non-linear mode interaction is an explanation for the time-dependent phenomena observed in KIC~5950759, as predicted by \citet{Dziembowski1985a} and \citet{Moskalik1985}. 

We obtained and analysed low-resolution spectroscopy of KIC~5950759 using the {\sc GSSP} software package \citep{Tkachenko2015b}. The determined effective temperature places KIC~5950759 within the predicted and observed instability region for \dsct stars (see \citealt{Murphy2019a}), albeit somewhat closer to the red edge but nonetheless within the typical parameter space of high-amplitude \dsct stars \citep{Petersen1996, McNamara2000b}. On the other hand, our spectroscopic analysis was unable to accurately constrain the rotational velocity of KIC~5950759 owing to the total broadening being smaller than the instrumental broadening. Also, the low resolution of the spectrum increased the uncertainties associated with determining the surface gravity and metallicity of KIC~5950759, but the best-fitting values provided by {\sc GSSP} are given in Table~\ref{table: parameters}. We compared previous determinations of the luminosity of KIC~5950759 by \citet{Murphy2019a} to new calculations based on Gaia EDR3 parameters \citep{Gaia2020a*} and conclude that KIC~5950759 is a single star as opposed to a binary.

Using the observed pulsation mode frequencies from the \Kepler mission data, the spectroscopic constraints on $T_{\rm eff}$ and metallicity, the Gaia luminosity, and the normalised instability parameters for the fundamental and first overtone, we determined the most likely masses and ages of KIC~5950759 using forward seismic modelling based on Petersen diagrams. Our results are summarised in Table~\ref{table: models}. We find that the best-fitting seismic models considering all the available constraints exist within two branches of the sub-solar metallicity regime (i.e. $Z \in (0.005, 0.006)\cup(0.010, 0.011)$). Our spectroscopic metallicity favours the lower-metallicity range in our models (i.e. $Z \in (0.005, 0.006)$, which further strengthens the conclusion that KIC~5950759 is near the TAMS. Multi-epoch high resolution spectroscopy that effectively samples the dominant pulsation cycle of KIC~5950759 is needed to accurately determine its atmospheric parameters. Large amplitude pulsations induce significant changes in the effective temperature and radius of a star, and such behaviour is not captured within our analysis of a single low resolution spectrum.

Most interestingly, the relatively large increasing period changes in the fundamental and first overtone radial modes combined with our modelling results indicates that such an observed period change is too large to be caused directly by stellar evolution. This is in agreement with observations of \dsct stars from the ground (see e.g. \citealt{Breger1998c}). Our modelling results indicate that KIC~5950759 is close to, if not just past, the TAMS. Despite this being the parameter space where period changes caused by evolution are predicted to be maximal, the observed amplitude and phase modulation of KIC~5950759 is too large to be directly caused by evolution. Given that the high-amplitude pulsations of HADS stars are sufficient to drive a periodic change in the stellar radius of order a factor two every $\simeq2$~hr, we posit that the observed amplitude and phase modulation is the manifestation of the star's structure responding to a highly non-linear driving mechanism. Post-main sequence \dsct or post-main sequence HADS stars are not uncommon, but since HADS stars themselves make up only a small minority of all A and F stars, this only lends further evidence that KIC~5950759 is a unique object within all the known \dsct stars observed by the \Kepler mission.

In the future, we expect ongoing space photometry missions, such as TESS \citep{Ricker2015}, to find additional interesting HADS stars in, or close to, relatively short lived stages of stellar evolution. Currently, it is difficult to study the evolutionary changes in \dsct stars from their pulsations in real time as they occur on such long time scales compared to observations. Moreover, non-linear mode interactions causing significant amplitude and phase modulation seemingly dominate their temporal behaviour \citep{Breger2000d, Breger2014, Bowman2014, Bowman2016a}. However, for relatively short-lived phases of evolution, such as the overall contraction phase and the immediate post-main sequence, with a large enough sample of \dsct stars observed by TESS, we should expect to increase this sample size to a non-negligible number of candidates. Ultimately, a large enough sample of pulsating stars with similar temporal properties to KIC~5950759 will allow us to probe non-linear pulsations and their effect on stellar structure in real time across a range of stellar masses and metallicities, and determine why such large period changes are present in so many \dsct stars, but not in all.

%%%%%%%%%%%%%%%%%%%%%%%%%%%%%%%%%%%%%%%%%%%%%%%%%%

\section*{Acknowledgements}

DMB gratefully acknowledges a senior postdoctoral fellowship from the Research Foundation Flanders (FWO) with grant agreement No.~1286521N. JDD acknowledges support from the Polish National Science Center (NCN), grant no. 2018/29/B/ST9/02803. DLH acknowledges the Science and Technology Facilities Council (STFC) via grant ST/M000877/1. SJM is a DECRA fellow supported by the Australian Research Council, grant number DE180101104. This work was supported by the Australian Research Council, an Australian Government Research Training Program (RTP) scholarship, and the Danish National Research Foundation (Grant DNRF106) through its funding for the Stellar Astrophysics Centre (SAC). The research leading to these results has received funding from the European Research Council (ERC) under the European Union's Horizon 2020 research and innovation programme (grant agreement No.~670519: MAMSIE), from the KU Leuven Research Council (grant C16/18/005: PARADISE), from the Research Foundation Flanders (FWO) under grant agreements G0H5416N (ERC Runner Up Project), as well as from the BELgian federal Science Policy Office (BELSPO) through PRODEX grant PLATO.

The \Kepler data presented in this paper were obtained from the Mikulski Archive for Space Telescopes (MAST) at the Space Telescope Science Institute (STScI), which is operated by the Association of Universities for Research in Astronomy, Inc., under NASA contract NAS5-26555. Support to MAST for these data is provided by the NASA Office of Space Science via grant NAG5-7584 and by other grants and contracts. Funding for the Kepler/K2 mission was provided by NASA’s Science Mission Directorate. The Gaia data in this paper come from the European Space Agency mission Gaia, processed by the Gaia Data Processing and Analysis Consortium (DPAC). Funding for the DPAC has been provided by national institutions, in particular the institutions participating in the Gaia Multilateral Agreement. The WASP project is funded and operated by Queen's University Belfast, the Universities of Keele, St. Andrews and Leicester, the Open University, the Isaac Newton Group, the Instituto de Astrof{\'i}sica de Canarias, the South African Astronomical Observatory, and by STFC. 

This research made use of the \texttt{Starlink} software \citep{STARLINK_2014}, which is currently supported by the East Asian Observatory. This research has made use of the SIMBAD database, operated at CDS, Strasbourg, France; the SAO/NASA Astrophysics Data System; and the VizieR catalogue access tool, CDS, Strasbourg, France. This research has made use of community-developed core \texttt{PYTHON} packages for astronomy (\texttt{ASTROPY}; \citealt{Astropy_2013, Astropy_2018}), the \texttt{PYTHON} library for publication quality graphics (\texttt{MATPLOTLIB}; \citealt{Matplotlib_2007}), \texttt{Numpy} \citep{Numpy_2011}, and \texttt{SciPy} \citep{SciPy_2020}.

\section*{Data availability}
The \Kepler data underlying this article are available from the MAST archive (\url{http://archive.stsci.edu/kepler/}). All other data underlying this article are available from the corresponding author upon reasonable request.

%%%%%%%%%%%%%%%%%%%%%%%%%%%%%%%%%%%%%%%%%%%%%%%%%%



\bibliographystyle{mnras}
\bibliography{/Users/dominic/Documents/Research/Bibliography/master_bib.bib}

%%%%%%%%%%%%%%%%%%%%%%%%%%%%%%%%%%%%%%%%%%%%%%%%%%

%%%%%%%%%%%%%%%%% APPENDICES %%%%%%%%%%%%%%%%%%%%%

\appendix

\section{Frequency lists}
\label{appendix: freq lists}

The additional independent non-radial pulsation mode frequencies identified in the LC and SC \Kepler data of KIC~5950759 are provided in Tables~\ref{table: residual LC list} and \ref{table: residual SC list}, respectively.

% - - - - - % - - - - - % - - - - - % - - - - - %
\begin{table}
\centering
\caption{Additional independent frequencies extracted from the 4-yr LC \Kepler data of KIC~5950759 (i.e. after $\nu_1$ and $\nu_2$ and all their significant harmonics and combinations have been removed). Pulsation phases were calculated with respect to the time zero-point $t_0 = 2455688.770$~BJD (i.e. the midpoint of the 4-yr LC light curve).}
\begin{tabular}{r r r}
\hline
\multicolumn{1}{c}{Frequency} & \multicolumn{1}{c}{Amplitude} & \multicolumn{1}{c}{Phase} \\
\multicolumn{1}{c}{(d$^{-1}$)} & \multicolumn{1}{c}{(mmag)} & \multicolumn{1}{c}{(rad)} \\
\hline
$0.31925 \pm 0.00004$	&	$0.284 \pm 0.030$	&	$1.99 \pm 0.11$	\\
$16.00000 \pm 0.00006$	&	$0.202 \pm 0.030$	&	$2.03 \pm 0.15$	\\
$18.45070 \pm 0.00009$	&	$0.125 \pm 0.030$	&	$-0.17 \pm 0.24$	\\
$18.96828 \pm 0.00008$	&	$0.149 \pm 0.030$	&	$1.47 \pm 0.20$	\\
$20.34427 \pm 0.00005$	&	$0.229 \pm 0.030$	&	$2.24 \pm 0.13$	\\
$21.15701 \pm 0.00005$	&	$0.211 \pm 0.030$	&	$2.25 \pm 0.14$	\\
$21.49656 \pm 0.00004$	&	$0.291 \pm 0.030$	&	$1.69 \pm 0.10$	\\
$22.62858 \pm 0.00003$	&	$0.442 \pm 0.030$	&	$-2.44 \pm 0.07$	\\
$23.70270 \pm 0.00006$	&	$0.187 \pm 0.030$	&	$-0.97 \pm 0.16$	\\
$23.82188 \pm 0.00006$	&	$0.192 \pm 0.030$	&	$0.06 \pm 0.16$	\\
$24.67230 \pm 0.00003$	&	$0.380 \pm 0.030$	&	$-1.32 \pm 0.08$	\\
$25.36702 \pm 0.00004$	&	$0.295 \pm 0.030$	&	$-1.57 \pm 0.10$	\\
\hline
\end{tabular}
\label{table: residual LC list}
\end{table}
% - - - - - % - - - - - % - - - - - % - - - - - %

% - - - - - % - - - - - % - - - - - % - - - - - %
\begin{table}
\centering
\caption{Additional independent frequencies extracted from the residual 31-d SC \Kepler data of KIC~5950759 (i.e. after $\nu_1$ and $\nu_2$ and all their significant harmonics and combinations have been removed). Pulsation phases were calculated with respect to the time zero-point $t_0 = 2455200.895$~BJD (i.e. the midpoint of the 31-d SC light curve).}
\begin{tabular}{r r r}
\hline
\multicolumn{1}{c}{Frequency} & \multicolumn{1}{c}{Amplitude} & \multicolumn{1}{c}{Phase} \\
\multicolumn{1}{c}{(d$^{-1}$)} & \multicolumn{1}{c}{(mmag)} & \multicolumn{1}{c}{(rad)} \\
\hline
$15.9977 \pm 0.0019$	&	$0.42 \pm 0.05$	&	$1.46 \pm 0.11$	\\
$18.4513 \pm 0.0018$	&	$0.47 \pm 0.05$	&	$1.63 \pm 0.10$	\\
$18.9699 \pm 0.0018$	&	$0.45 \pm 0.05$	&	$0.91 \pm 0.10$	\\
$20.3421 \pm 0.0014$	&	$0.58 \pm 0.05$	&	$-0.62 \pm 0.08$	\\
$21.1629 \pm 0.0017$	&	$0.48 \pm 0.05$	&	$1.95 \pm 0.10$	\\
$21.4957 \pm 0.0007$	&	$1.19 \pm 0.05$	&	$-2.47 \pm 0.04$	\\
$22.6277 \pm 0.0012$	&	$0.69 \pm 0.05$	&	$-2.24 \pm 0.07$	\\
$23.7050 \pm 0.0021$	&	$0.39 \pm 0.05$	&	$-1.18 \pm 0.12$	\\
$23.8250 \pm 0.0012$	&	$0.71 \pm 0.05$	&	$-0.84 \pm 0.06$	\\
$24.6676 \pm 0.0009$	&	$0.88 \pm 0.05$	&	$-1.03 \pm 0.05$	\\
$25.3680 \pm 0.0016$	&	$0.51 \pm 0.05$	&	$-0.75 \pm 0.09$	\\
$28.3737 \pm 0.0015$	&	$0.56 \pm 0.05$	&	$0.57 \pm 0.08$	\\
$33.6469 \pm 0.0013$	&	$0.64 \pm 0.05$	&	$2.83 \pm 0.07$	\\
\hline
\end{tabular}
\label{table: residual SC list}
\end{table}
% - - - - - % - - - - - % - - - - - % - - - - - %

%%%%%%%%%%%%%%%%%%%%%%%%%%%%%%%%%%%%%%%%%%%%%%%%%%

% Don't change these lines
\bsp	% typesetting comment
\label{lastpage}
\end{document}